\documentclass[twocolumn,showpacs,aps,superscriptaddress]{revtex4-1}
\usepackage{amsmath}
\usepackage[switch, modulo]{lineno}
\usepackage{graphicx}
\usepackage{longtable}
\usepackage{ltablex}
\include{def}

\def\be{\begin{equation}}
\def\ee{\end{equation}}

\newcommand{\beq}{\begin{equation}}\newcommand{\eeq}[1]{\label{#1}
\end{equation}}\newcommand{\beqar}{\begin{eqnarray}}\newcommand{\eeqar}[1]
{\label{#1}
\end{eqnarray}}\newcommand{ \mpt }{\langle p_{T}
\rangle}\newcommand{\bmath}{\begin{displaymath}}\newcommand{\emath}{\end{displaymath}}\newcommand{\bitem}{\begin{itemize}}\newcommand{\eitem}{\end{itemize}}
 
\begin{document}

\title{\Large \bf Multiplicity dependent $p_T$ distributions of identified 
particles in pp collisions at 7 TeV within HIJING/$B$$\bar{B}$ v2.0 model}

\newcommand{\mcgill}{Physics Department, McGill University, Montreal, Canada, H3A 2T8}
\newcommand{\ifin}{National Institute for Physics and Nuclear Engineering-Horia~Hulubei\\
Hadron Physics Department\\ 
R-077125, Bucharest, Romania}

\author{~V.~Topor~Pop} \affiliation{\mcgill}\affiliation{\ifin}
\author{~M.~Petrovici} 
\email{mpetro@nipne.ro}
\affiliation{\ifin}

\date{\today}

\begin{abstract}
Effects of strong longitudinal color fields (SLCF) on the 
identified (anti)particle transverse momentum ($p_T$) distributions 
in $ pp$ collision at $\sqrt{s}$ = 7 TeV are investigated 
within the framework of the {\small HIJING/B\=B v2.0} model.
The comparison with the experiment is performed 
in terms of the correlation between mean 
transverse momentum ($\mpt$) and multiplicity (N$_{ch}^*$) 
of charged particles at central rapidity, as well as
the ratios of the $p_T$ distributions to the one corresponding to 
the minimum bias (MB) pp collisions at the
same energy, each of them normalized to the corresponding 
charged particle density, for high multiplicity 
\mbox{(HM, N$_{ch}>100$)} 
and low multiplicity \mbox{(LM, N$_{ch}<100$)} class of events.
The theoretical calculations show that an increase 
of the strength of color fields (as characterized by the 
effective values of the string tension $\kappa$), from 
$\kappa$ = 2 GeV/fm to $\kappa$ = 5 GeV/fm from LM to 
HM class of events, respectively, 
lead to a ratio at low and intermediate $p_T$ 
({\it i.e.}, \mbox{$ 1$GeV/{\it c} $< p_{T} < 6 $ GeV/{\it c}}) consistent
with recent data obtained at LHC by the ALICE Collaboration. 
These results point out to the necessity of introducing 
a multiplicity (or energy density) dependence 
for the effective value of the string tension.
Moreover, the string tension $\kappa$ = 5 GeV/fm, describing the $p_T$ 
spectra of ID (anti)particle in pp collisions at $\sqrt{s}$ = 7 TeV for high
charged particle (HM) multiplicity event classes, 
has the same value as the one used in 
describing the $p_T$ spectra in central Pb - Pb collisions 
at $\sqrt{s_{\rm NN}}$ = 2.76 TeV. 
Therefore, we can conclude that at the LHC energies the global features 
of the interactions could be mostly determined by the properties of 
the initial chromoelectric flux tubes, while the system size may 
play a minor role.  

\end{abstract}


\maketitle

\section{Introduction}

Relativistic and ultra-relativistic heavy-ion experimental data 
evidenced global features such as flow, baryon-meson anomaly, 
(multi)strange enhancement, jet quenching which 
support the interpretation within theoretical (phenomenological) models  
as originating from a deconfined, strongly interacting thermalised phased, coined Quark-Gluon Plasma (sQGP). 
In contrast, no similar effects
were observed in proton-proton ($pp$) and proton-nucleus ($p-A$) collisions, these results being considered of interest only as reference data for nucleus-nucleus ($A - A$) collisions.
Features reminiscent from heavy-ion phenomenology have been 
recently evidenced
in such reactions at the LHC energies, {\it i.e.}, long range near side ridge in particle correlations 
\cite{Khachatryan:2010gv,CMS:2012qk,Abelev:2012ola,Aad:2013fja}, collective flow \cite{Andrei:2014vaa,Aad:2015gqa,Witek:2017dyn} 
or strangeness enhancement \cite{ALICE:2017jyt} observed in high charged particle multiplicity events.
The nature of these similarities is still an open question. 
Do they originate from a deconfined phase following a hydrodynamic 
evolution like in nucleus-nucleus ($A - A$) collisions or are they 
a consequence of the initial state dynamics manifested in the 
final state observables 
\cite{Andrei:2014vaa,Schlichting:2016sqo,Fischer:2016zzs,Bierlich:2017vhg} ? 
Most probable the two processes coexist, with a dense thermalised central 
core and an outer corona. Such a picture is successfully implemented 
in the EPOS model ({\underline E}nergy sharing {\underline P}arton based theory with {\underline O}ff-shell remnants and ladder {\underline S}plitting)\cite{Pierog:2013ria,Werner:2013tya,Werner:2014xoa}. The core-corona interplay 
in the light flavor hadron production for Pb - Pb collisions at $\sqrt{s_{NN}}$ = 2.76 TeV was recently discussed in Ref. \cite{Petrovici:2017izo}.
Therefore, the study of of $pp$, $p - A$ and $A - A$ collisions as a function of 
charged particle multiplicity has gathered recently much attention 
both, experimentally and theoretically. 

The non-perturbative particle creation mechanisms in strong external fields play an important role from
$e^{+}e^{-}$ pair creation in quantum electrodynamics (QED) \cite{schwinger}, up to pair 
creation of fermions and bosons in strong non-Abelian fields 
\cite{Biro:1984cf,Bial,Gyulassy:1986jq,Meri,Tanji:2008ku,
Ruffini:2009hg,Labun:2008re,Cardoso:2011cs,Nayak:2005pf,
Nayak:2005yv,Levai:2008wf,Levai:2009mn,Levai:2011zza,Levai:2011zz}.
In high-energy heavy-ion collisions, strong color fields are expected 
to be produced between the partons of the projectile and target.
Particle production in high energy $pp$ and
$A - A$ collisions can be described within chromoelectric flux tube
({\em strings}) models \cite{Andersson:1983ia,Wang:1991hta,Wang:1991vx}.

In a string fragmentation phenomenology, it has been proposed
that the observed strong enhancement of strange particle
production transverse momentum distribution in nuclear collisions
could be naturally explained via strong longitudinal 
color field effects (SLCF)~\cite{Biro:1984cf,Bial,Gyulassy:1986jq,Meri}.
Recently, an extension
of color Glass Condensate (CGC) theory has proposed a more detailed
dynamical model
of color ropes ``GLASMA'' \cite{larry_2009,McLerran:2008qi,Kharzeev:2006zm}.

Strong longitudinal fields 
(flux tubes, effective strings) decay into new ones by 
 quark anti-quark ($q\bar{q}$ ) or 
diquark anti-diquark (qq-$\overline{\rm qq}$) pair 
production before hadronization. Due to confinement, the color 
of these strings is restricted to a small area in transverse space
\cite{Cardoso:2011cs}.
With increasing energy of the colliding particles, the number of
strings grows and they start to overlap, producing clusters. 
This is the origin of the energy density dependence of particle production \cite{Braun:2012kn}. The effect of modifying the string tension
due to local density has been studied in Monte Carlo models, which are
used primarily for heavy-ion collisions 
\cite{Sorge:1992ej,Bass:1998ca,Bleicher:1999xi,Soff:2000ae,Soff:2002bn,Lin:2004en}. In the Partons String Models (PSM) string fusion and percolation effects 
on strangeness and heavy flavor production have also been  
discussed in Refs.~\cite{Braun:2015eoa,Bautista:2017gds,Merino:2009py,Pajares:2010vm}. A similar model with string fusion into color ropes is considered in the 
Dipole evolution in Impact Parameter Space and rapidity
({\small DIPSY}) \cite{Bierlich:2017vhg,Bierlich:2014xba,Bierlich:2015rha}. 
String collective effects were also introduced 
in a multi-pomeron exchange model  to improve the production of 
hadrons in  $pp$ collisions at the LHC energies
\cite{Feofilov:2017udk,Bodnya:2014pta,armesto2008}.

Heavy Ion Jet Interacting ({\small HIJING}) type models, 
 \cite{Wang:1991hta,Wang:1991vx},
 {\small HIJING2.0} \cite{Deng:2010mv,Deng:2010xg}
and {\small HIJING/B\=B} v2.0 
\cite{Pop:2004dq,Pop:2005uz,ToporPop:2007hb,Pop:2009sd,ToporPop:2010qz,Pop:2013ypa,ToporPop:2011wk,Barnafoldi:2011px,Pop:2012ug,Albacete:2013ei,Albacete:2016veq}, were developed in order to 
explain the hadron production 
in $pp$, $p - A$ and $A - A$ collisions.   
These approaches are based on a two-component geometrical 
model of mini-jet production and soft interaction and incorporate 
nuclear effects such as {\em shadowing} and {\em jet quenching}, via final state jet
medium interaction.  
{\small HIJING/B\=B} v2.0 model \cite{ToporPop:2007hb,ToporPop:2010qz}
includes new dynamical effects associated with
long range coherent fields ({\it i.e}, strong longitudinal color fields, SLCF),
via baryon junctions and loops \cite{Pop:2005uz,ripka_lnp2004}.  
At RHIC it was shown  \cite{Pop:2004dq,Pop:2005uz,ToporPop:2007hb}
that the dynamics of strangeness production 
deviates from calculations based on Schwinger-like
estimates for homogeneous and constant color fields \cite{schwinger}, 
pointing to a possible 
contribution of fluctuations of transient SLCF.
These fields are rather similar to those which could appear in a 
 {\em GLASMA} \cite{McLerran:2008qi} at initial stage of the collisions.
The typical field strength of SLCF at ultra-relativistic energies,
in a scenario with QGP phase transition, 
was estimated to be about 5-12 GeV/fm \cite{Magas:2002ge}.

Global observables and identified particle (ID) data,  
including (multi)strange particles production in  {\it p}{\it p} 
\cite{Pop:2013ypa,ToporPop:2010qz,Pop:2012ug}
{\it p} - Pb \cite{Barnafoldi:2011px,Albacete:2013ei,Albacete:2016veq}
and Pb - Pb collisions \cite{ToporPop:2011wk} at the LHC energies
were successfully described by
{\small HIJING/B\=B v2.0} model.
However, correlations among different measurable quantities in 
multi-particle production offer a better way to constrain the models.
In this paper we extend our study to identified particle ({\it i.e.}, $\pi$, $K$, $p$, $\Lambda$, $\Xi$, $\Omega$ and their anti-particle) produced in 
small collision systems. We will perform a detailed analysis of 
correlations between average transverse momentum 
$ \mpt $ and charged particles multiplicity (N$_{ch}^*$) and 
for the ratio of double differential cross sections normalized to the charged particle densities ($dN_{ch}/d\eta$) versus multiplicity, {\it i.e.},

\begin{equation}
{\large R_{mb}}\,\, (cen) \,\,=\,\, \left(\frac{\frac{d^2N}{dydp_T}}{\langle \frac{dN_{ch}}{d\eta}\rangle }\right)_i^{cen} / \left(\frac{\frac{d^2N}{dydp_T}}{\langle \frac{dN_{ch}}{d\eta}\rangle }\right)_i^{ppMB}
\label{eq:rmb} 
\end{equation}
\noindent
where i = identified particle in $pp$ collisions, "cen'' stand for 
multiplicity event classes. We will consider 
high multiplicity (HM; N$_{ch} > 100$), 
and low multiplicity (LM; N$_{ch} < 100$ ) classes.
MB stand for minimum bias events.
The charged particle densities $dN_{ch}/d\eta$ are integrated values at mid-pseudo-rapidity $|\eta| < 0.5$ for that class of events.
The $p_T$ distributions of ID particle were recently measured  in $pp$ collisions  at $\sqrt{s}$ = 7 TeV for different multiplicity classes of events by ALICE Collaboration 
~\cite{Bianchi:2016szl,DerradideSouza:2016kfn,Vislavicius:2017lfr,Longpaper}.

\section{HIJING/B\=B v2.0 model.}
The {\small HIJING1.0} model has been discussed in detail in our
previous papers \cite{ToporPop:2010qz,Pop:2013ypa,Pop:2012ug}.
Here we briefly summarize the main assumptions and parameters determined as in \cite{ToporPop:2010qz}.
   
The production rate for a quark pair ($q\bar{q}$) 
per unit volume for a uniform chromoelectric flux tube with field ({\it E}) being: 
 
\begin{equation}
\Gamma =\frac{\kappa^2}{4 \pi^3} 
{\text {exp}}\left(-\frac{\pi\,m_{q}^2}{\kappa}\right),
\label{eq:Gamma}
\end{equation}
\noindent
\cite{Gyulassy:1986jq,Cohen:2008wz,Hebenstreit:2008ae}, 
strong chromoelectric fields are required, 
$\kappa/m_{\rm q}^2\,\,>$ 1, for a significant production rate.
Consequently,  the production rate of heavy quark pair $Q\bar{Q}$
is suppressed by a factor $\gamma_{Q\bar{Q}}$ \cite{Cohen:2008wz}:

\begin{equation}
\gamma_{Q\bar{Q}} = \frac{\Gamma_{Q\bar{Q}}}{\Gamma_{q\bar{q}}} =
{\text {exp}} \left(-\frac{\pi(m_{Q}^2-m_q^2)}{\kappa} \right), 
\label{eq:gammaQ}
\end{equation}

The suppression factors are calculated for 
$Q={\rm qq}$ (diquark), $Q = s$ (strange), $Q = c$ (charm), or $Q = b$ (bottom) ($q$ means $u$ or $d$ quark).
The quark masses used in the present paper are:   
$m_s=0.12$ GeV, $m_{\rm c}=1.27$ GeV, $m_b=4.16$ GeV \cite{pdg:2010},
and for di-quark $m_{\rm qq}$ = 0.45 GeV \cite{ripka:2005}.
The following effective masses
$M^{\rm eff}_{qq}$ = 0.5 GeV, 
$M^{\rm eff}_{s}$ = 0.28 GeV and $M^{\rm eff}_{c}$ = 1.27 GeV
have been considered.
For these values and a vacuum string tension $\kappa_0$ = 1 GeV/fm, Eq.~\ref{eq:gammaQ} gives the following suppression of heavier quark production pairs:
$u\bar{u}$ : $d\bar{d}$ : {\rm qq}$\overline{\rm qq}$ : $s\bar{s}$ : $c\bar{c}$ $\approx$ 1 : 1 : 0.02 : 0.3 : 10$^{-11}$ \cite{Pop:2012ug}. 
On the other hand,
if the effective string tension value $\kappa$ 
increases to  $\kappa = f_{\kappa} \kappa_0$ (with $ f_{\kappa} > 1$), as it is the case for a colour rope,
the value of $\gamma_{Q\bar{Q}}$ increases.
A similar increase of $\gamma_{Q\bar{Q}}$ is obtained if the
quark mass decreases from $m_{Q}$ to  $m_{Q}/\sqrt{f_{\kappa}}$.
It was shown \cite{ToporPop:2010qz} that such a dynamical mechanism gives better agreement with the measured strange meson/hyperon 
ratios at the Tevatron and at LHC energies.
It is also known that the $A - A$ collision data are reproduced 
using flux tubes with much larger string tension relative to
the 
fundamental string tension linking a mesonic quark-antiquark pair 
\cite{Biro:1984cf,Cardoso:2011cs}. 
As the initial energy densities produced in the collision, 
$\epsilon_{\rm ini}$, is proportional to mean field values $<E^2>$
\cite{Cardoso:2011cs},
and $\kappa= e_{\rm eff} E$, 
$\epsilon_{\rm ini} \propto \kappa^2$. 
Based on Bjorken approach the $\epsilon_{\rm ini}$ is proportional with 
charged particle density at mid-rapidity. Therefore,  
$ \kappa^2 \propto (dN_{\rm ch}/d\eta)_{\eta =0} $ and 
$\kappa \propto Q_{sat,p}$, similar with CGC model, as it was
discussed in Ref.~\cite{Pop:2013ypa}. The energy dependence 
of the charged particle density at mid rapidity in $pp$ collisions
up to the LHC energies was described using a power law dependence:   
\begin{equation} 
\kappa(s)= \kappa_{0} \,\,(s/s_{0})^{0.04}\,\,{\rm GeV/fm},
\label{eq:kappa_sup}
\end{equation}
consistent with the value  
deduced in CGC model for $Q_{sat,p}$ \cite{McLerran:2010ex}.
  
Following equation~\ref{eq:kappa_sup}, at $\sqrt{s}$ = 0.2 TeV the effective
string tension value is
$\kappa$ = 1.5 GeV/fm while at $\sqrt{s}$ = 7 TeV 
$\kappa$ = 2.0 GeV/fm.
Previous papers \cite{Pop:2005uz,ToporPop:2007hb,Pop:2009sd,ToporPop:2010qz,ToporPop:2011wk,Pop:2012ug} presented the dependence of 
different observables on the string tension values.
The phenomenological parametrisation Eq.~\ref{eq:kappa_sup},
is supported  by the experimental results on charged particle densities at mid-rapidity $(dN_{\rm ch}/d\eta)_{\eta =0}$. 
Within the error bars the  $\sqrt{(dN_{\rm ch}/d\eta)_{\eta =0}}$ shows a power law $s^{0.05 }$ dependence
for inelastic $pp$  and $s^{0.055}$ dependence
for non-single diffractive events ~\cite{Aamodt:2010pb,ALICE:2012xs}.
In $A - A$ collisions the effective string tension value 
could also increase due to in-medium effects~\cite{ToporPop:2011wk},
or as a function of centrality. 
This increase is considered in our phenomenology by an analogy 
with CGC model, i.e.: 
$\kappa(s,A) \propto Q_{\rm sat,A} (s,A) \propto  Q_{\rm sat,p} (s) A^{1/6}$.
Therefore, in the present analysis for $A - A$ collisions
we used $\kappa$ = $\kappa(s,A)$:
\begin{equation}
\kappa(s,A)_{\text LHC} = \kappa(s) A^{0.167}
= \kappa_{0} \,\,(s/s_{0})^{0.04} A^{0.167}\,\,{\rm GeV/fm},
\label{eq:kapsA} 
\end{equation}

Eq.~\ref{eq:kapsA} gives $\kappa(s,A)_{\text LHC} \approx 5$ GeV/fm, 
for Pb - Pb collisions at $\sqrt{s_{NN}}$ = 2.76 TeV.  
The suppression factor $\gamma_{Q\bar{Q}}$,  
approach unity in Pb - Pb collisions at $\sqrt{s_{NN}}$ = 2.76 TeV, 
for $\kappa \geq 5$ GeV/fm. 
The mean effective values of the string tension 
$\kappa(s)$ for $pp$ collisions (Eq.~\ref{eq:kappa_sup}) 
and $\kappa(s,A)$ for Pb - Pb  collisions (Eq.~\ref{eq:kapsA})
are used in the present calculations. As a concequence
the various suppression factors 
and the intrinsic (primordial) 
transverse momentum $k_T$ increase
\cite{ToporPop:2010qz,Pop:2013ypa}.
We would like to mention also that for a better description of the baryon/meson anomaly evidenced
at RHIC and LHC energies we introduced a specific J\=J loops,  
(for details see Refs.~\cite{ToporPop:2011wk,Pop:2012ug}).
The absolute yield of charged particles, $dN_{\rm ch}/d\eta$, is sensitive
to the low $p_T < \, 2$ GeV/{\it c} 
non-perturbative hadronization
dynamics. This was considered based on LUND \cite{Andersson:1986gw} 
string JETSET \cite{Bengtsson:1987kr} fragmentation 
constrained by lower energy $ee, ep, pp$ data.  
The hard pQCD contribution is estimated 
in {\small HIJING/B\=B v2.0} using {\small PYTHIA}
\cite{Sjostrand:2006za} subroutines. Details on shadowing and jet quenching are given in Ref.~\cite{Pop:2013ypa}. 
The main advantage of {\small HIJING/B\=B v2.0}  over {\small PYTHIA 6.4}
resides in the SLCF colour rope effects 
that arise from longitudinal fields amplified by the random walk 
in color space of the high x valence partons in {\it A} - {\it A} collisions.  
A broad fluctuation spectrum of the effective string tension could be induced by this random walk.
The present work is focussed on the effect 
of a larger effective value $\kappa > $ 1 GeV/fm
on the production of identified particles 
measured in Pb - Pb, $p$ - Pb and $pp$ collisions at LHC energies.
While the present approach is based on the time-independent strength of colour field, in 
reality the production of $Q\bar{Q}$ pairs is a far-from-equilibrium,
time and space dependent complex, phenomenon.
Therefore, the influence of time dependent fluctuations can't be addressed within the present approach. 

\section{Numerical Results and Discussion}

\subsection{The average transverse momentum $ \mpt $ versus 
N$_{ch}$ correlations}

The {\small HIJING/B\=B v2.0} model 
predicts many experimental observables (charged hadron pseudo-rapidity distributions, transverse momentum spectra, identified particle spectra,
baryon-to-meson ratios) using the above values for the effective 
string tension, $\kappa$ (see Sec. II) 
~\cite{ToporPop:2010qz,Pop:2013ypa,Barnafoldi:2011px,Pop:2012ug,Albacete:2013ei}.

\begin{figure*}[tbh!]
\centering
\includegraphics[width=0.48\textwidth,height=5.5cm]{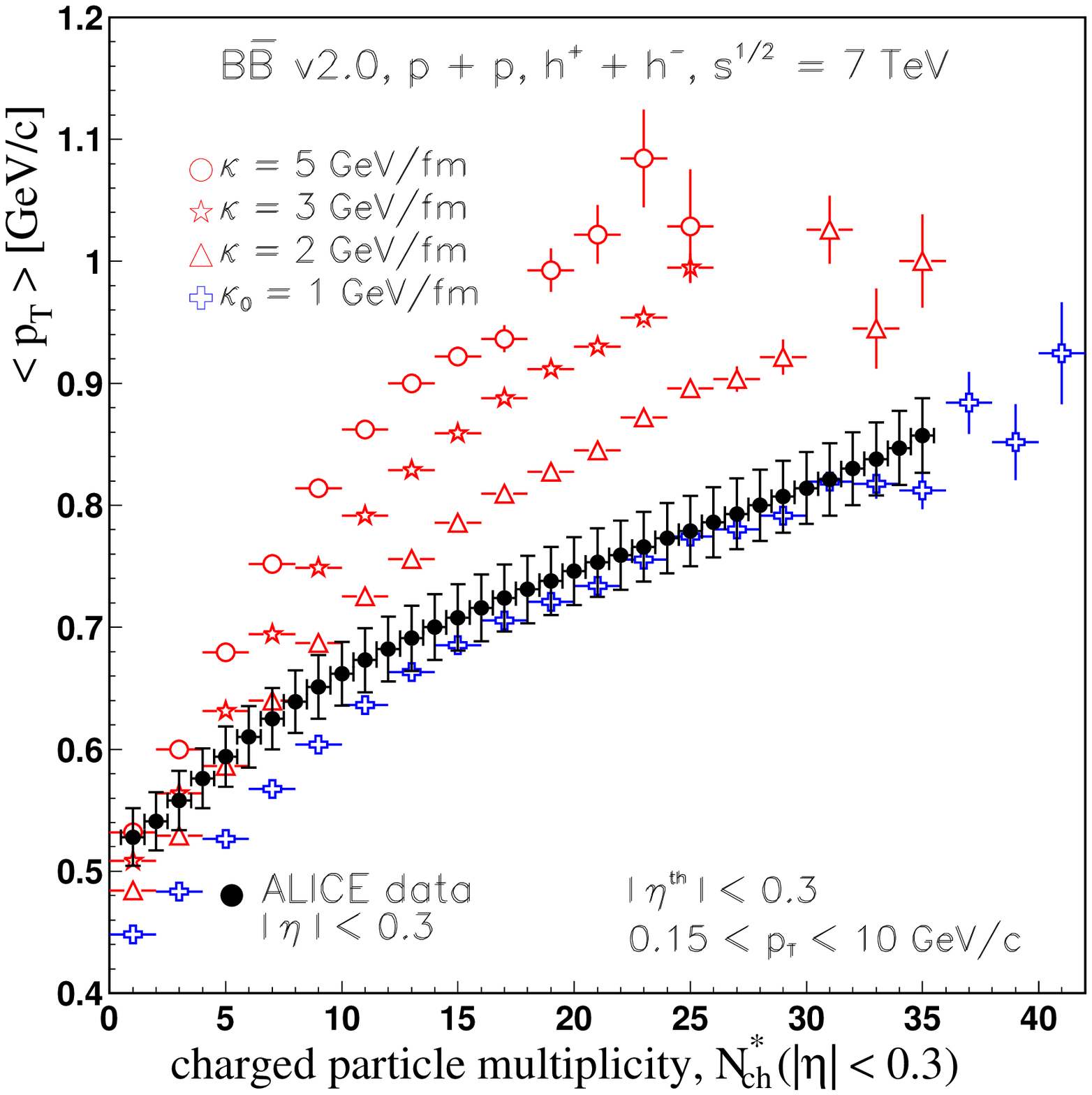}
\includegraphics[width=0.48\textwidth,height=5.5cm]{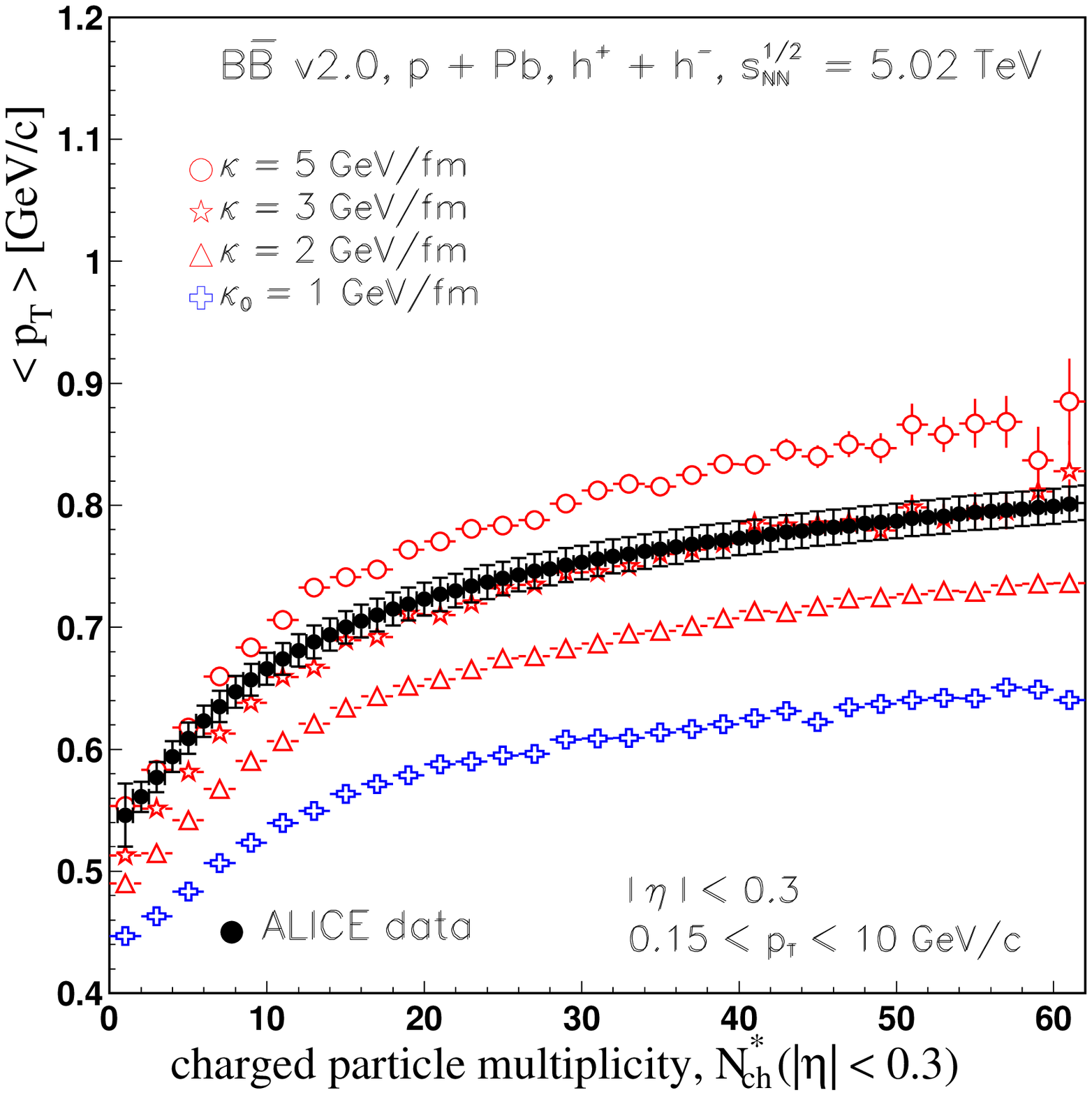}
\vskip 0.5cm
\caption[pp 7tev meanpt vs nch id part ks5] 
{\small Open symbols-{\small HIJING/B\=B v2.0} predictions for
the average transverse momentum ($ \mpt $) of charged particles    
as a function of multiplicity at mid-pseudo-rapidity
$N_{ch}^*$. Left panel-$pp$ collisions at $\sqrt{s}$ = 7 TeV for
$ 0.15 < p_T < 10$ GeV/{\it c} and mid-pseudo-rapidity $|\eta| < 0.3$;
Right panel-$p$ - Pb  collisions at $\sqrt{s_{\rm NN}}$ = 5.02 TeV for
$ 0.15 < p_T < 10$ GeV/{\it c} and mid-rapidity $|\eta| < 0.3$. 
The theoretical results are obtained for different effective string tensions increasing from $\kappa = $ 1 GeV/fm (default) up to $\kappa = $ 5 GeV/fm. The ALICE data (filled circles) are from Ref.~\cite{Abelev:2013bla}.
The errors represent systematic uncertainties on $\mpt $. The statistical errors are negligible.
\label{fig:mpt_alch_pp7tev_ppb5tev}
}
\end{figure*}

The ALICE Collaboration has reported measurements of the 
average transverse momentum $ \mpt$ versus charged particles 
$N_{ch}^*$ at central rapidity in $pp$ at $\sqrt{s}$ = 7 TeV , $p$ -Pb 
at  $\sqrt{s_{\rm NN}}$ = 5.02 TeV, and Pb - Pb collisions  at $\sqrt{s_{\rm NN}}$ = 2.76 TeV~\cite{Abelev:2013bla}. The analysis range was restricted to a transverse momentum  $ 0.15 < p_T < 10$  GeV/{\it c} and to a 
mid-pseudo-rapidity range $|\eta| < 0.3$.
Figure \ref{fig:mpt_alch_pp7tev_ppb5tev} shows the results obtained 
with {\small HIJING/B\=B v2.0} model (open symbols) for $pp$ collisions at $\sqrt{s}$ = 7 TeV (left panel) and  $p$ - Pb at  $\sqrt{s_{\rm NN}}$ = 5.02 TeV (right panel).
As we can see in Fig. \ref{fig:mpt_alch_pp7tev_ppb5tev} a continuous increase 
of $\mpt$ with $N_{ch}^*$ is observed for both reactions.
Therefore, to calculate the correlation  $ \mpt_{N_{ch}}$ vs $N_{ch}^*$ ,
we first investigate in a model of hadronizing strings if 
the above increase could be attributed to the effects of SLCF and
the results are given for different strength of color fields quantified 
by an effective value of the string tension from
$\kappa$ = 1 GeV/fm (default value) up to $\kappa$ = 5 GeV/fm.
As we could remark, the calculations with the default value $\kappa$ = 1 GeV/fm, describe better the $pp$ data. 
An alternative explanation of the  increase of  $\mpt$ with $N_{ch}^*$ 
should be naturally given in the context of the 
fragmentation of multiple minijets 
embedded in {\small HIJING} type models \cite{Wang:1991hta} and was discussed in the early 90s for 
$p\bar{p}$ collisions at  $\sqrt{s}$ = 1.8 TeV \cite{Bial,Wang:1991vx,Meri}.
The large multiplicity events are dominated by multiple minijets  
while low multiplicity events are dominated by those of no jet production.
Few partons are enough to explain the increase of $\mpt$ with $N_{ch}^*$.
We may also conclude that these correlations in $pp$ collisions at  $\sqrt{s}$ = 7 TeV are not sensitive to the soft fragmentation region, where we expect that SLCF effects are dominant. 
In contrast, for  $p$ - Pb collisions the theoretical calculations compared to data~\cite{Abelev:2013bla} in Fig.~\ref{fig:mpt_alch_pp7tev_ppb5tev} (right panel) show  better agreement if the value of $\kappa$ is increased from  $\kappa$ = 1 GeV/fm to $\kappa$ = 3 GeV/fm.

\begin{figure*} [htb]
\centering
\includegraphics[width=0.48\textwidth,height=5.5cm]{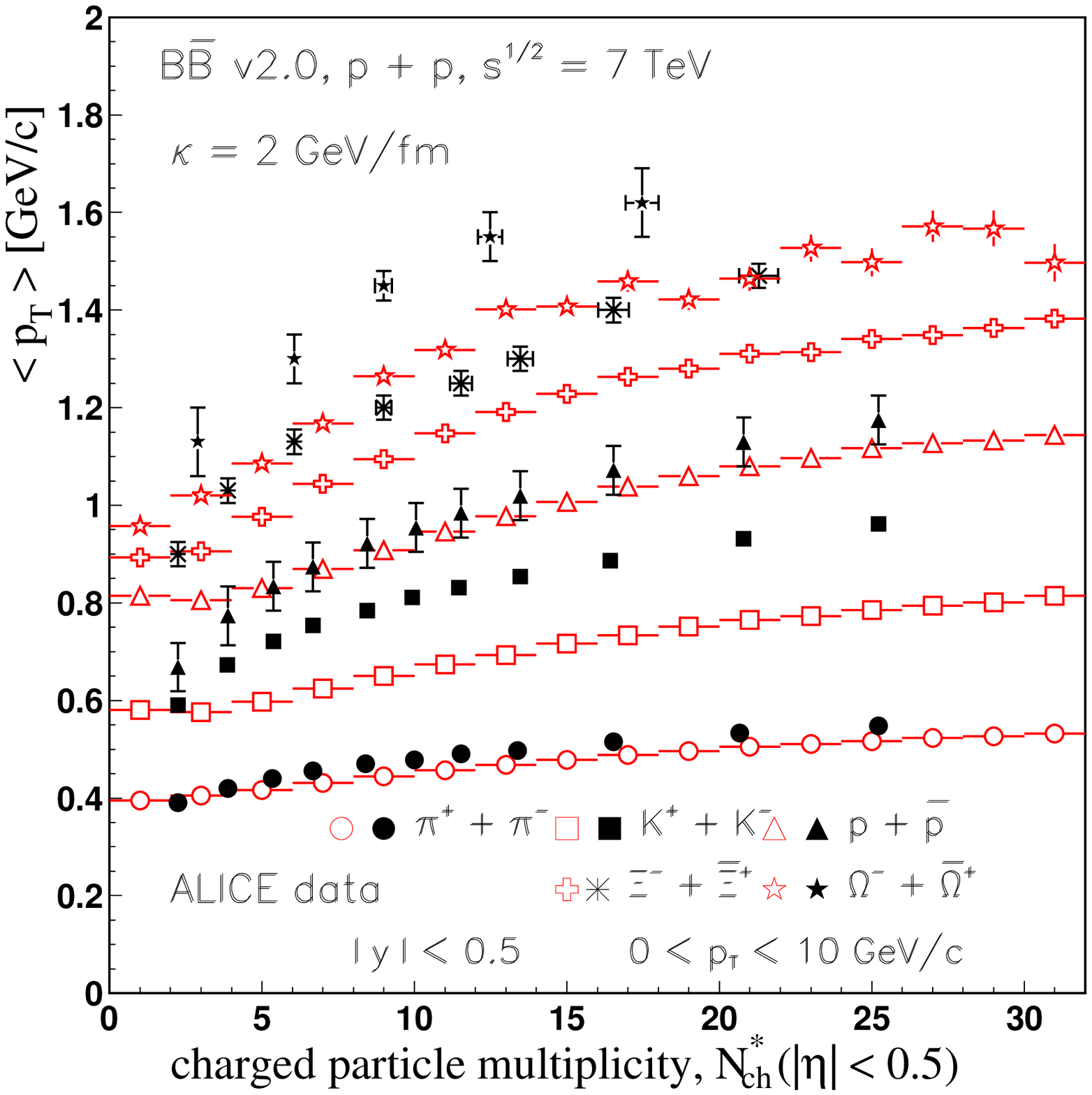}
\includegraphics[width=0.48\textwidth,height=5.5cm]{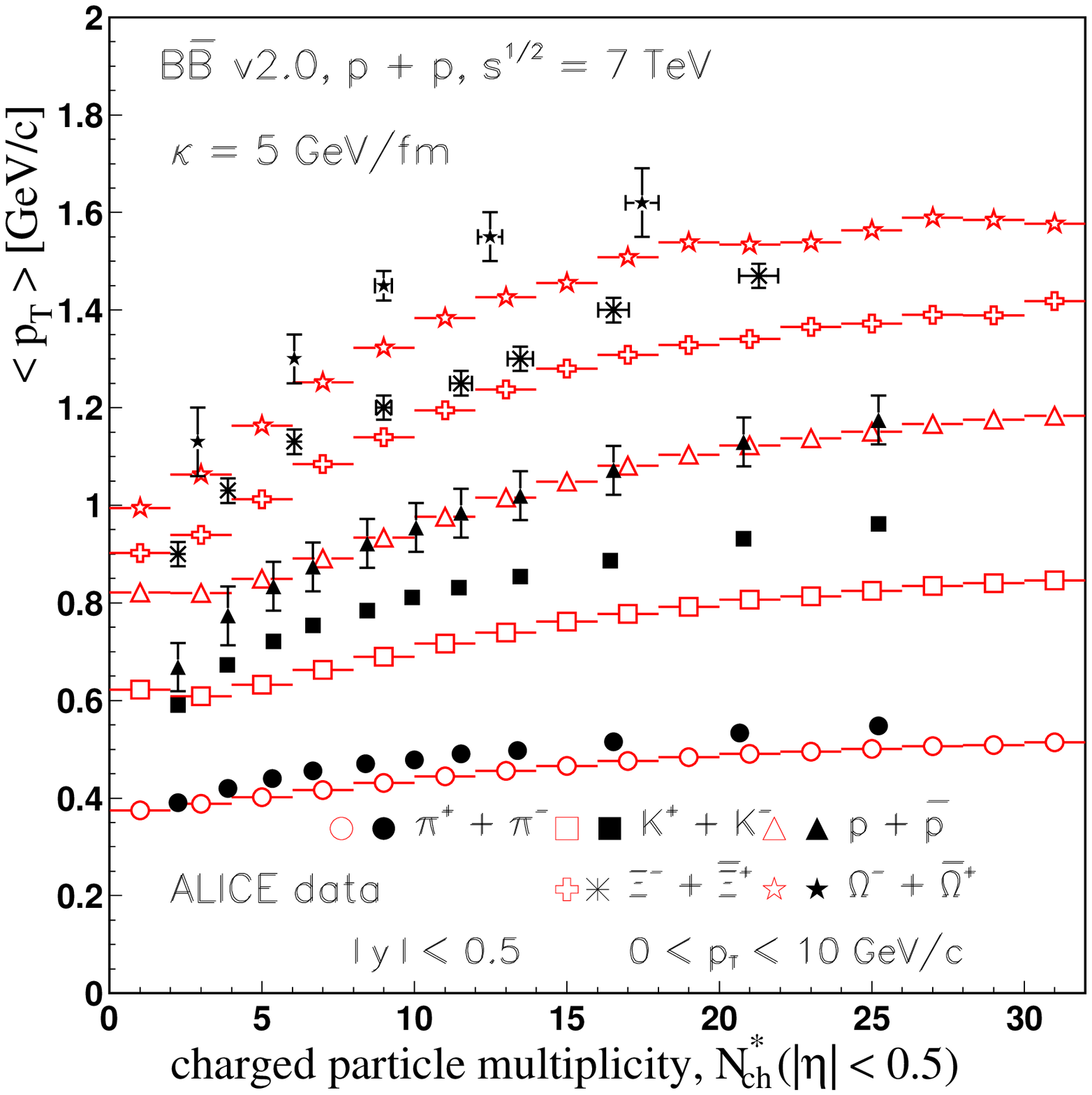}
\vskip 0.5cm
\caption[pp 7tev meanpt vs nch id part ks5] 
{\small Open symbols-{\small HIJING/B\=B v2.0} predictions for
the average transverse momentum ($ \mpt $) of 
identified particle for $ 0 < p_T < 10$ GeV/{\it c} and 
mid-rapidity $|y| < 0.5$ as 
function of charged particle multiplicity, $N_{ch}^*$ in $pp$ collisions at 
$\sqrt{s}$ = 7 TeV. The results are obtained with an effective string tension value, $\kappa = $ 2 GeV/fm (left side) and $\kappa = $ 5 GeV/fm (right side). For clarity we do not include the results for $\Lambda$.
The ALICE preliminary data (filled symbols) are from Ref.~\cite{Bianchi:2016szl,DerradideSouza:2016kfn, Longpaper}. 
Only statistical error bars are shown.
\label{fig:mpt_pp7tev_id_ks2_ks5}
}
\end{figure*}

We will study now the effect of an enhanced value of the effective 
string tension $\kappa$ on the correlation of $\mpt$ versus  
$N_{ch}^*$ for ID particle in $pp$ and $p$ - Pb collisions 
at  $\sqrt{s}$ = 7 TeV, and $\sqrt{s_{\rm NN}}$ = 5.02 TeV, respectively.
Shown in Fig.~\ref{fig:mpt_pp7tev_id_ks2_ks5} are our theoretical calculations 
(open symbols) in comparison with data~\cite{Bianchi:2016szl,DerradideSouza:2016kfn,Longpaper}  on the $ \mpt $  
of $\pi^+ +\pi^-$, $K^+ + K^-$, $p+\bar{p}$, $\Xi^- + {\bar{\Xi}}^+$, and $\Omega^-+ {\bar{\Omega}}^+$ 
for $ 0 < p_T < 10$  GeV/{\it c} and mid-rapidity $|y| < 0.5$ versus charged particle multiplicity   $N_{ch}^*$  (selected in the $|\eta| < 0.5$ range)
for $pp$ collisions at $\sqrt{s}$=7 TeV. 
The results (open symbols) are given for two values of the effective string tension 
 $\kappa = $ 2 GeV/fm (left panel) and $\kappa = $ 5 GeV/fm (right panel). 
The data show an increase of $ \mpt $ with increased multiplicity
and with the particle mass, facts 
fairly well described by the model. 
Note, that for clarity we did not include here the results for 
$\Lambda + \bar{\Lambda}$. Since the mass difference between 
lambda and proton is very small, the results are almost the same~\cite{Bianchi:2016szl}.
\noindent
The $ \mpt $ increases with increasing multiplicity 
as the effect of the strong longitudinal color field embedded in our model. A modified string fragmentation using 
$\kappa = $ 2 GeV/fm increase the production rate 
for heavier particles. 
Moreover an increase of the width of the primordial (intrinsic) transverse momentum ($k_T$) distribution  
from the default value of the Gaussian ($\sigma_{q}$ = $\sigma_{\rm qq}$ = 0.350 GeV/{\it c})  to larger values for the (anti)quark ($\sigma_{q}'' = \sqrt{\kappa/\kappa_0} \cdot \sigma_{q}$)
and  (anti)diquark ($\sigma_{\rm {qq}}'' = \sqrt{\kappa/\kappa_0}
\cdot {\rm f} \cdot \sigma_{{\rm qq}}$), where f = 3 ~\cite{ToporPop:2010qz,Pop:2013ypa}, 
contribute also to an increases of the  heavier particles production rate. 
This provides a consistent  
evidence that modified fragmentation obtained by an enhanced $\kappa$ from the default value 
$\kappa $ = 1 GeV/fm and minijet production as implemented 
in  {\small HIJING/B\=B v2.0} model lead to a fairly good description 
of these observables. 
However, the model give only partial agreement of  $ \mpt $ values for
ID particle at high multiplicity.
The model describes well $ \mpt $ of $\pi^+ +\pi^-$,  $p+\bar{p}$, $\Lambda + \bar{\Lambda}$, but  the results strongly underestimate the $ \mpt $ of (multi)strange particles as  $K^+ + K^-$, and $\Xi^- + {\bar{\Xi}}^+$, and $\Omega^- + {\bar{\Omega}}^+$.
We studied if one can find a scenario that would give a larger enhancement of the $ \mpt $ of (multi)strange particles. We consider the effect of a further increase of the string tension to $\kappa = $ 5 GeV/fm and the results are presented in Fig.~\ref{fig:mpt_pp7tev_id_ks2_ks5} (right panel). 

Note that a value $\kappa \approx 5 \kappa_0$ GeV/fm is also supported by 
the calculations at finite temperature ($T$) 
of potentials associated with a $q\bar{q}$ pair separated by a
distance $r$ \cite{Liao:2008vj}.
The finite temperature ($T$) form of the $q\bar{q}$ potential
has been calculated by means of lattice QCD \cite{Kaczmarek:2005ui}.
At finite temperature, there are two potentials associated
with a $q\bar{q}$ pair separated by a distance $r$:
the free energy $F(T,r)$ and potential energy $V(T,r)$.
The free and potential energies actually correspond to slow and fast
(relative) motion of the charges, respectively.
Infrared sensitive variables such as  string tension,
their derivatives with respect to $r$,
are very helpful to identify specific
degrees of freedom of the plasma. 
Since the confinement of color in non-Abelian theories is due to
the magnetic degree of freedom, the magnetic component is 
expected to be present in the plasma as well.
In the presence of the {\em chromo-magnetic scenario} it was shown that 
the effective string tension of the free energy $\kappa$ = $\kappa_F$ 
decreases with $T$, to near zero at critical temperature ($T_c$). 
In contrast, the effective string tension of the potential energy
(corresponding to a fast relative motion of the charges)
$\kappa$ = $\kappa_V$ remains nonzero below $\sim$$T = 1.3 \,T_c$
with a peak value at $T_c$ of about 5 times the vacuum tension
$\kappa_{0} $ ($\kappa_V$ = 5 GeV/fm)
\cite{Liao:2008vj,Shur}.

The above calculations for  $\kappa \approx 5 \kappa_0$ GeV/fm
result in only a modest increase of the $ \mpt $ of kaons ($K^+ + K^-$)
by 10-15 \% and a better description of  $ \mpt $  of multi-strange particles
($\Xi^- + {\bar{\Xi}}^+$ and $\Omega^- + {\bar{\Omega}}^+$) only at 
low multiplicity  ($N_{ch} < $ 15).
In our calculations the discrepancy obtained for $ \mpt $ of kaons does not 
appear to turn over for  $\kappa = $ 5 GeV/fm as expected.
This discrepancy may be related to the kaon enhancement 
reported first in Ref.~\cite{Alexopoulos:1990hn} at Tevatron energies and 
confirmed now at LHC energies \cite{Bianchi:2016szl,DerradideSouza:2016kfn,
Longpaper}.
Note, that new {\small PYTHIA8} model which includes 
a specific increase of the string tension values \cite{Fischer:2016zzs}, 
also could not describe better the  $ \mpt $ of kaons in $pp$ collisions at 
$\sqrt{s}$ = 7 TeV. Further analysis are necessary in order to draw a definite 
conclusion.  

\begin{figure*} [thb]
\centering
\includegraphics[width=0.48\textwidth,height=5.5cm]{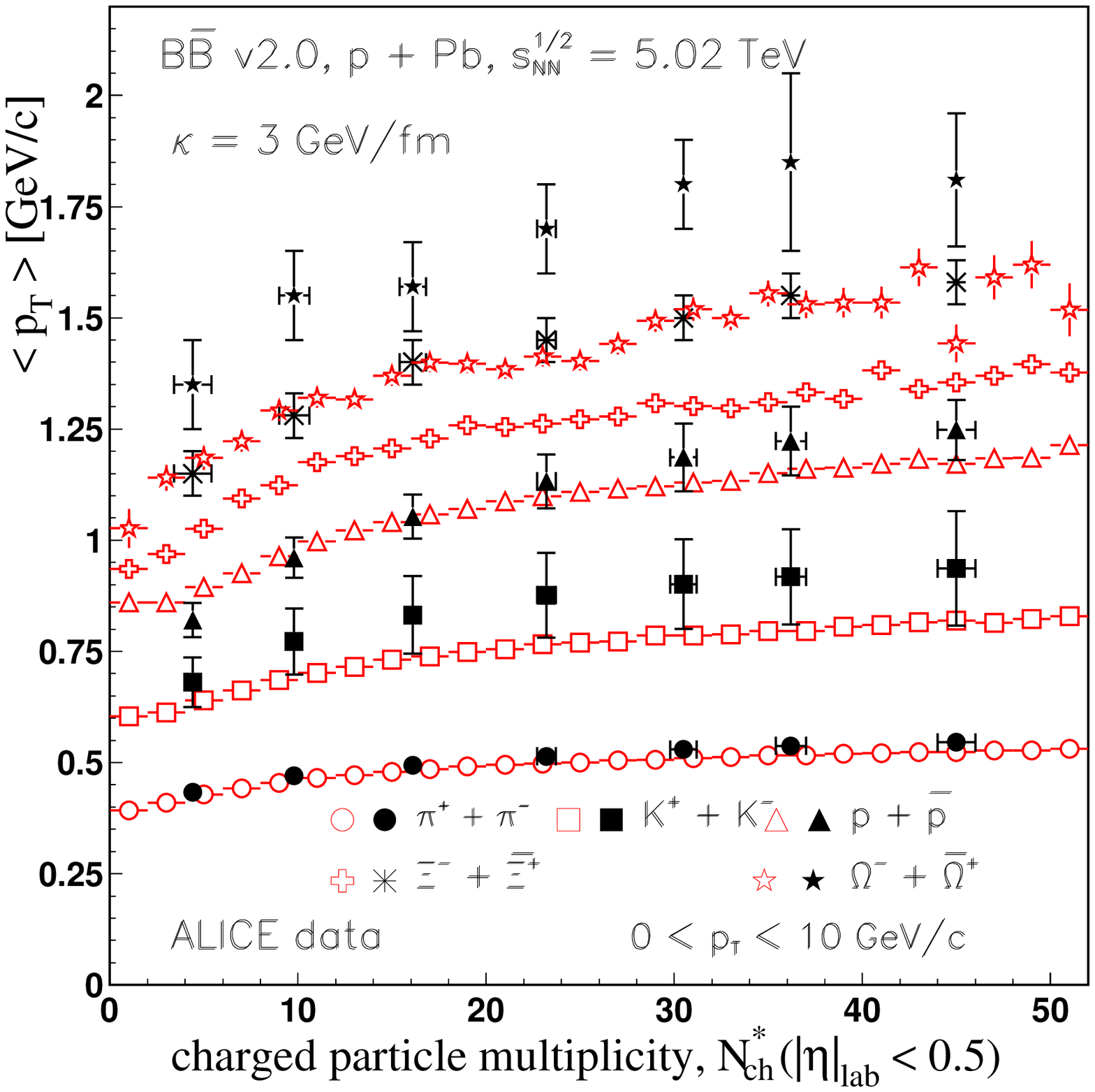}
\includegraphics[width=0.48\textwidth,height=5.5cm]{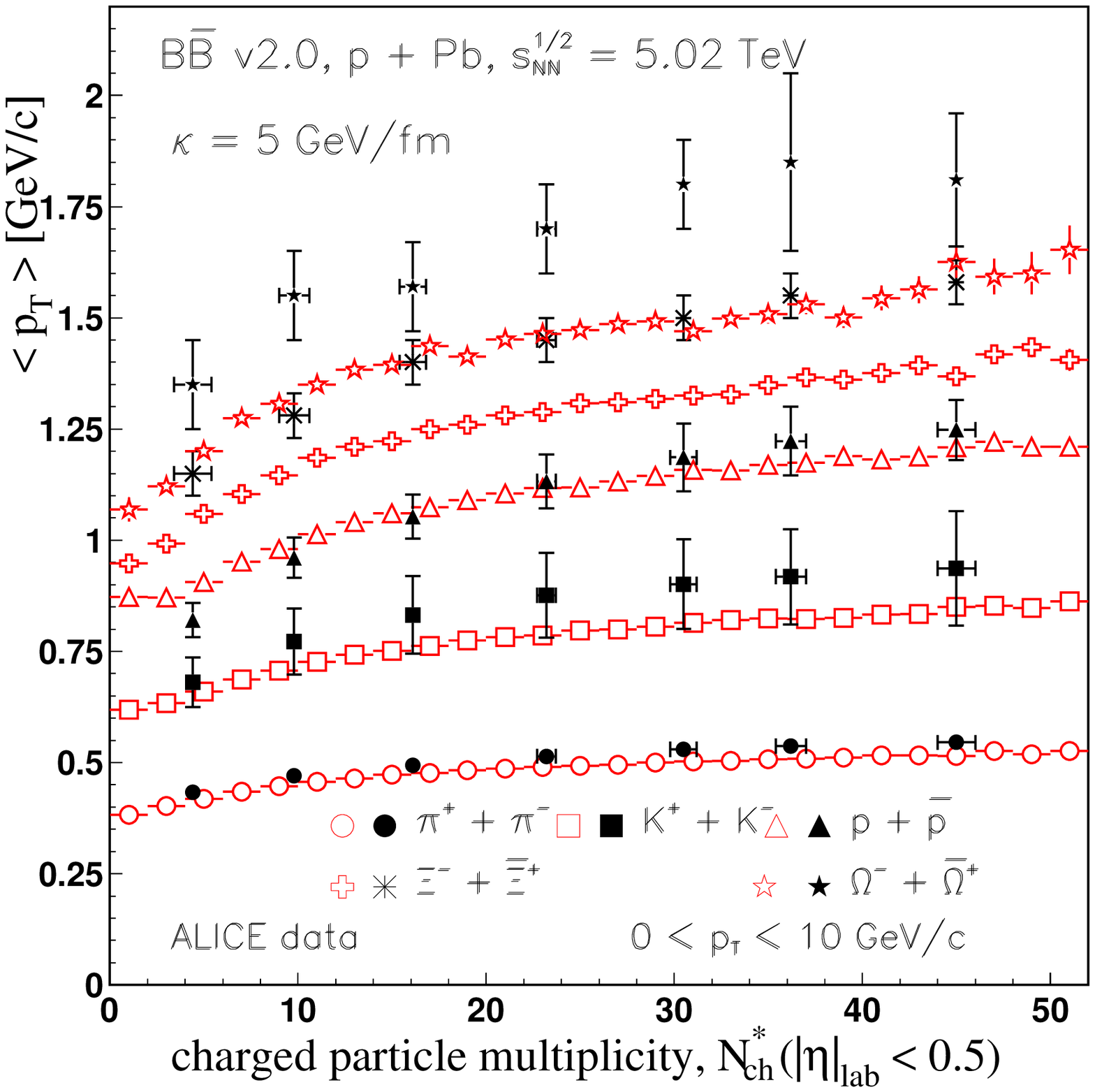}
\vskip 0.5cm
\caption[ppb 5tev meanpt vs nch id part ks3] 
{\small Open symbols - {\small HIJING/B\=B v2.0} predictions for
the average transverse momentum ($ \mpt $) of 
identified particle in the range $ 0 < p_T < 10$  GeV/{\it c} and mid-rapidity $0.0 < y_{cm} < 0.5$ as function of charged particle multiplicity, 
$N_{ch}^*$ in $p $ - Pb collisions at 
$\sqrt{s_{\rm NN}}$ = 5.02 TeV. The results (open symbols) are obtained with an effective string tension value, $\kappa = $ 3 GeV/fm (left side),and $\kappa = $ 5 GeV/fm (right side). For clarity we do not include the results for $\Lambda$.
The ALICE data (filled symbols) are from Ref.~\cite{Abelev:2013haa}.
Only statistical error bars are shown.
\label{fig:mpt_ppb5tev_id_ks3_ks5}
}
\end{figure*}

In the {\small HIJING/B\=B v2.0} model the collective behavior is a consequence of the confining strong color fields, resulting in an interaction between strings that is without diffusion or loss of energy \cite{Bierlich:2017vhg}.   
Therefore, for values of string tension between 
5 and 10 GeV/fm (the calculations are not included here) a saturation seems 
to set in, possibly as an effect
of energy and momentum conservation, as well as due to a saturation of 
strangeness suppression factors.
Similar conclusions could be drawn for $ \mpt $ of ID particles 
versus  charged particle multiplicity, 
$N_{ch}^*$ measured in $p $ - Pb collisions at 
$\sqrt{s_{\rm NN}}$ = 5.02 TeV. The results (open symbols) are obtained 
in the range $ 0 < p_T < 10$  GeV/{\it c} 
and mid-rapidity $0.0 < y_{cm} < 0.5$, and are shown in 
Fig.~\ref{fig:mpt_ppb5tev_id_ks3_ks5}.

\begin{figure*} [t!]
\centering
\includegraphics[width=0.32\textwidth,height=5cm]{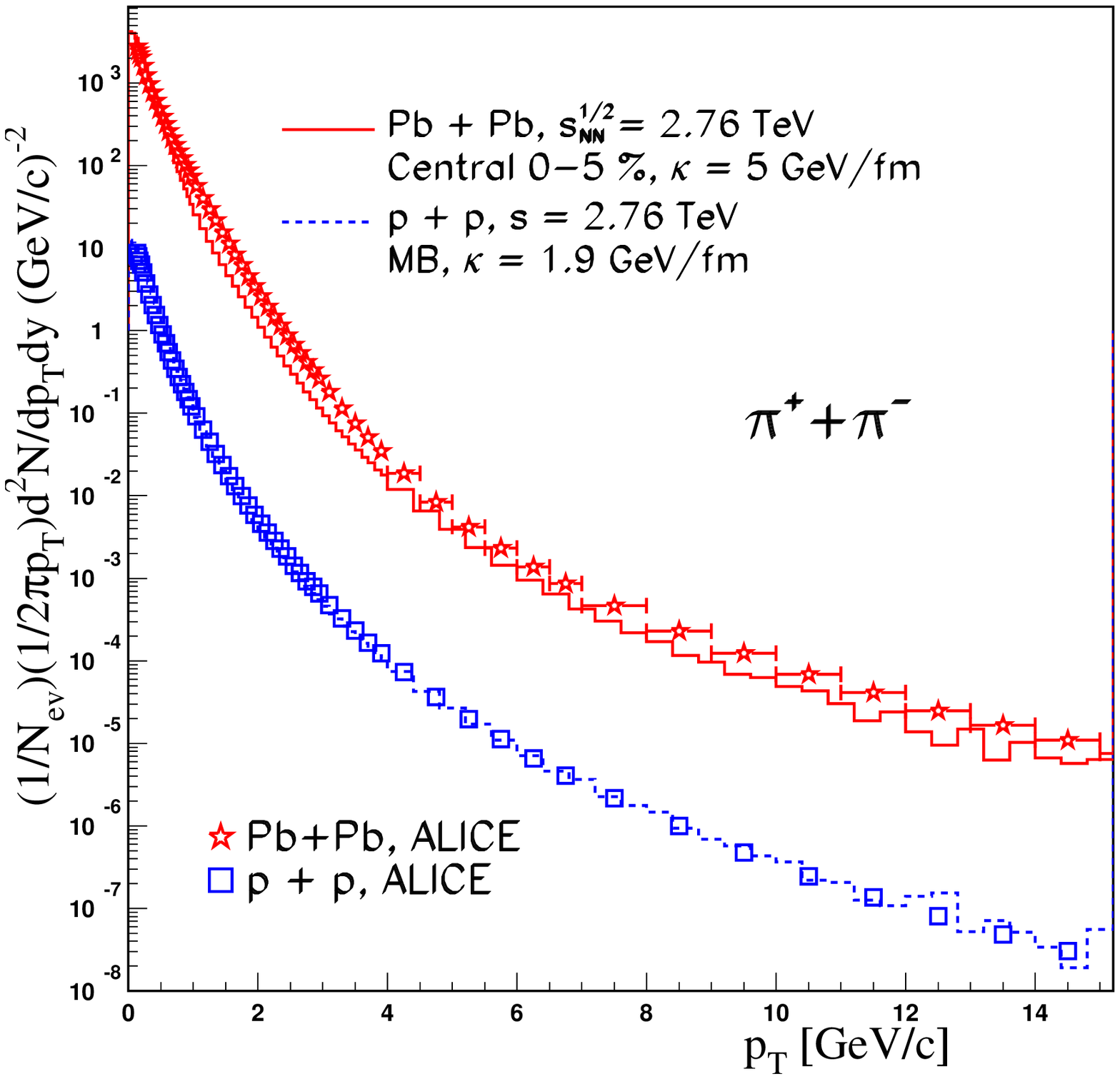}
\includegraphics[width=0.32\textwidth,height=5cm]{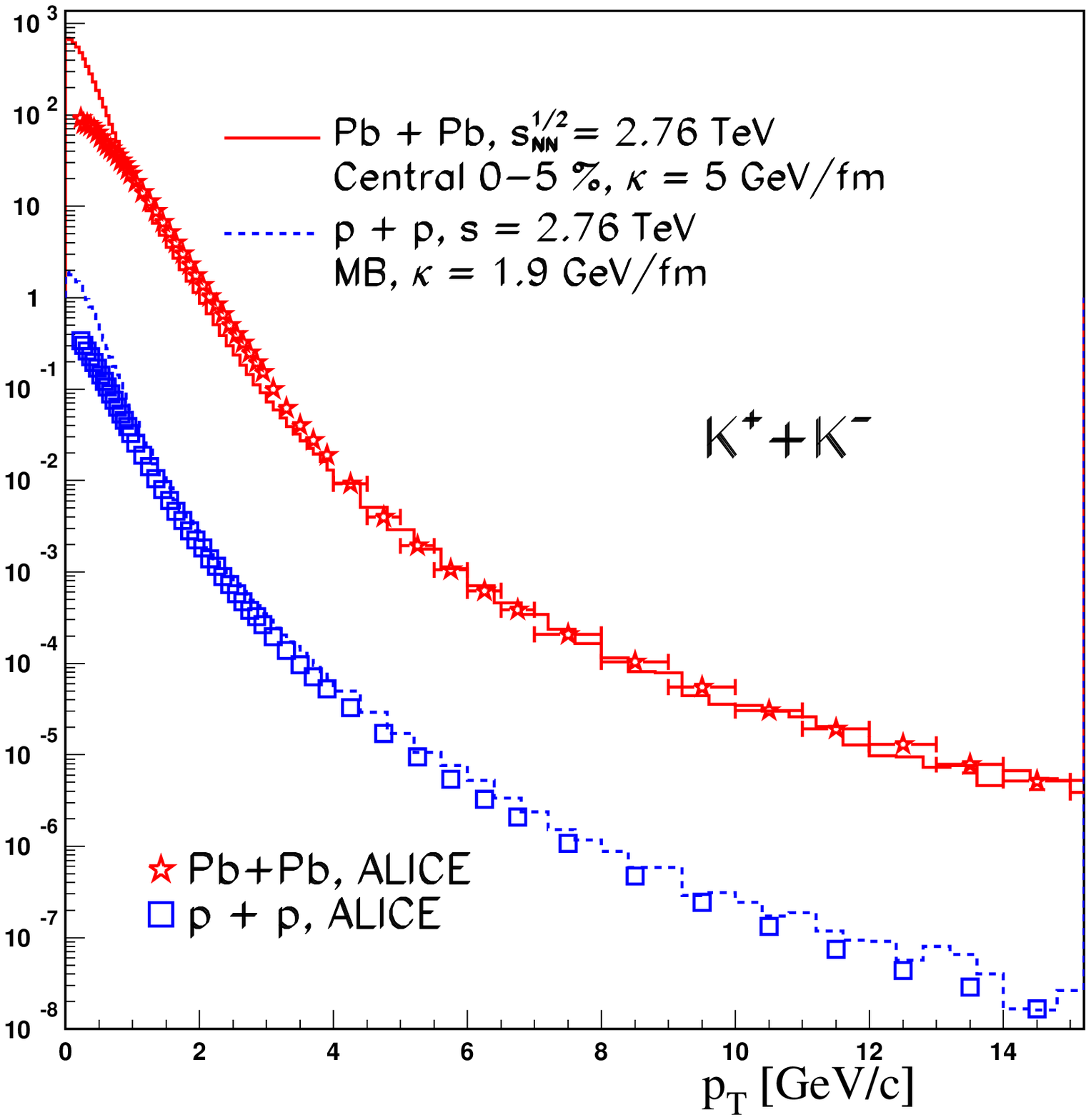}
\includegraphics[width=0.32\textwidth,height=5cm]{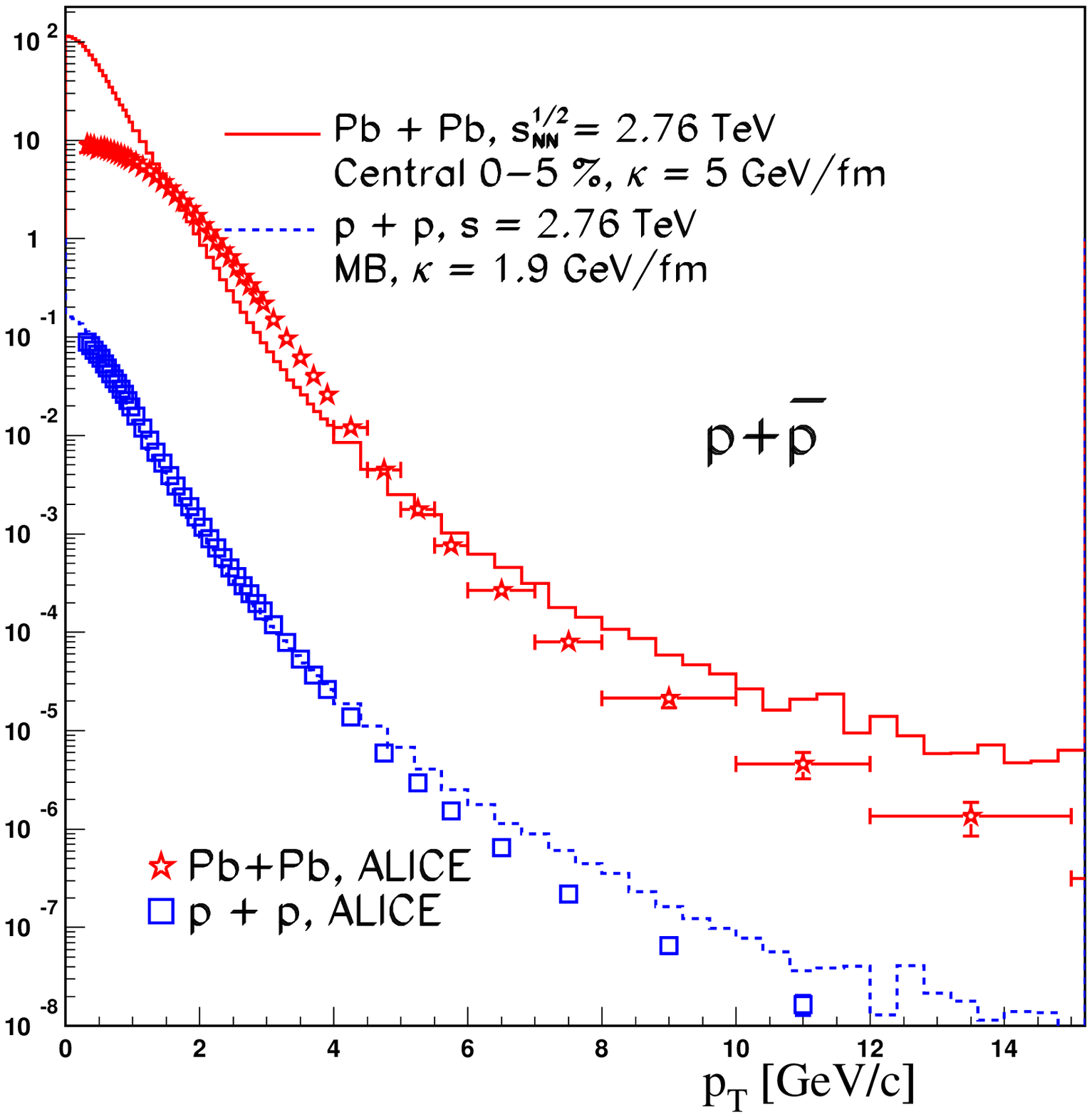}
\vskip 0.5cm
\caption[pbpb 2.76tev meanpt vs nch] 
{\small (color online) {\small HIJING/B\=B v2.0} predictions for
the invariant yields of identified particles in central 
Pb -Pb collisions (solid histograms)
and  $pp$ collisions (dashed histograms) at c.m. energy 2.76 TeV.
The results are obtained using $\kappa$ = 5 GeV/fm ($\kappa$ = 1.9 GeV/fm)
for Pb - Pb ($pp$) respectively.
The ALICE data are from Ref.~\cite{Abelev:2014laa}. The error include systematic uncertainties. 
\label{fig:pikp_pt_pbpb2.76tev}
}
\end{figure*}

Up to now, the microscopic origin of enhanced  
(multi)strange particles production 
is not known.  It is, therefore, a valid question whether small systems 
(high multiplicity $pp$ and $p$ - Pb )  exhibit 
any behavior of the kind observed in heavy-ion collisions. 
Bjorken suggested the possibility of deconfinement in $pp$ collisions
\cite{bjorken_1982}. Van Hove 
\cite{VanHove:1982vk} and Campanini \cite{Campanini:2011bj} suggested
that an anomalous behavior of average
transverse momentum ($ \mpt $) as a function of multiplicity could be
a signal for the occurrence of a phase transition in hadronic matter,
{\it i.e.}, formation of a {\em mini quark-gluon plasma} (mQGP).
The long range near side ridge in particle correlations observed in high multiplicity events 
\cite{Khachatryan:2010gv,CMS:2012qk,Abelev:2012ola,Aad:2013fja}, collective flow \cite{Andrei:2014vaa,Aad:2015gqa,Witek:2017dyn}
and strangeness enhancement \cite{ALICE:2017jyt} were evidenced in $pp$ collisions at the LHC energies and support such hypothesis.
However, a fundamental question remains, are such correlation of $\mpt$ vs $N_{ch}^*$ for ID particle in small 
systems ($pp$, $p$ - Pb collisions) of
collective origin, attributed 
to a hydrodynamic evolution like in 
Pb - Pb collisions, or they are a natural consequence due to initial state dynamics that show-up in the final state observables~\cite{Bierlich:2017vhg}?  
Collective hydrodynamic flow as a signature of sQGP is well 
established in Pb - Pb collisions
at LHC energies. Such conclusions can be drawn from measurements of
 the invariant transverse momentum spectra of identified particles in central 
Pb -Pb collisions   at $\sqrt{s_{\rm NN}}$ = 2.76 TeV.     
In Fig.~\ref{fig:pikp_pt_pbpb2.76tev} we consider the results 
for light identified charged hadrons in Pb - Pb collisions (solid histograms) in comparison with
those produced in  $pp$ collisions (dashed histograms) at 
$\sqrt{s}$ = 2.76 TeV.
The experimental data are from ALICE Collaboration Ref.~\cite{Abelev:2014laa}.
The calculations are performed taking an effective value of 
the string tension 
$\kappa$ with an energy and mass dependence as in Eq.~\ref{eq:kapsA}, {\it i.e.},
$\kappa(s,A)_{\text LHC} = \kappa(s) A^{0.167}
= \kappa_{0} \,\,(s/s_{0})^{0.04} A^{0.167}\,\,{\rm GeV/fm}$.
This formula leads to $\kappa(s,A)_{\text LHC} \approx 5$ GeV/fm, 
in Pb - Pb collisions at c.m. energy per nucleon $\sqrt{s_{\rm NN}}$ = 2.76 TeV.
In $pp$ collisions we consider only variation with energy, {\it i.e.}, 
$\kappa(s)= \kappa_{0} \,\,(s/s_{0})^{0.04}\,\,{\rm GeV/fm}$, which gives a
value of $\kappa$$\approx$1.9 GeV/fm.
The results obtained within our model show a partial agreement with data, 
since a large pressure in the initial state, leading to flow
especially for (anti)protons, 
is not considered in string fragmentation models.  

\subsection{Ratio of normalized transverse momentum  distributions}

The measured transverse momentum distributions for ID particles 
for different multiplicity bins have been recently
reported by ALICE Collaboration
in $pp$ collisions at $\sqrt{s}$ = 7 TeV 
\cite{Bianchi:2016szl,DerradideSouza:2016kfn,Vislavicius:2017lfr,
Longpaper}.
 The transverse momentum spectra of the identified hadrons (ID) were measured 
for several event multiplicity classes 
from the highest (class I) to the lowest (class X) multiplicity class,
corresponding to approximately 3.5 and 0.4 times the average value in the 
integrated sample ($<dN_{ch}/d\eta> ^{MB}\,\approx\, 6.0$), respectively.
In the experiment the multiplicity classes are defined based on the total charge deposited in the V0A and V0C detectors located at forward ($2.8 < \eta < 5.1$) 
and backward ($-3.7 < \eta<-1.7$) pseudorapidity regions, respectively.
The event multiplicity estimator is taken to be the sum of V0A and V0C signals denoted as V0M.
The average charged particle density ($ <dN_{ch}^{exp}/d\eta>$), is estimated within each such multiplicity class by the average of the
tracks distribution in the region  $|\eta| < 0.5$.

Based on these spectra and minimum bias results we will study here  
the ratio of double differential cross sections normalized to the charged particle densities $dN_{ch}^{th}/d\eta$ versus multiplicity, {\it i.e.}, 
the ratio $R_{mb}$ defined in Eq.~\ref{eq:rmb}. 

\begin{figure*} [th]
\centering
\includegraphics[width=8.3 cm,height=8.3 cm]{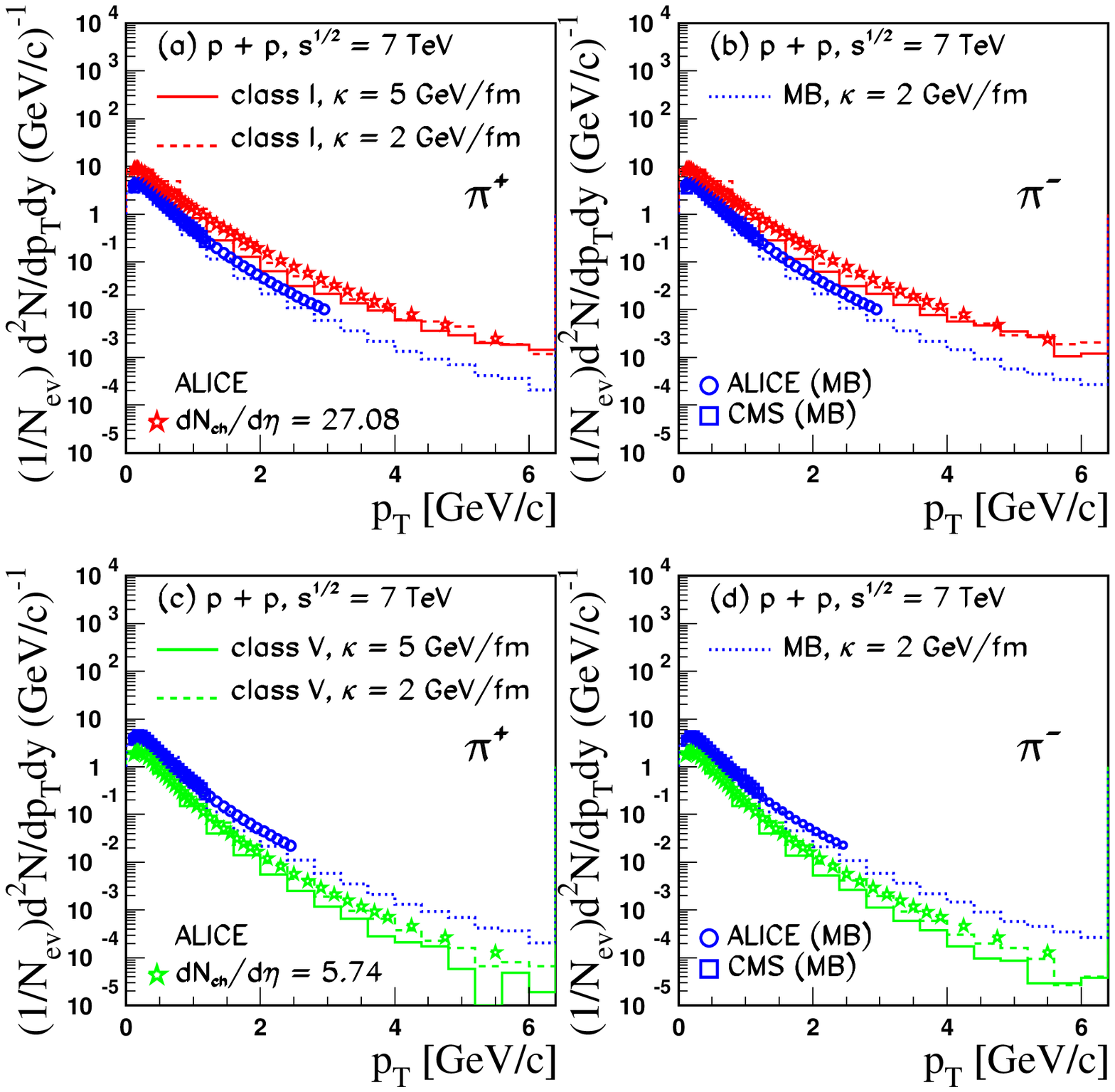}
\includegraphics[width=0.5\textwidth,height=4.0cm]{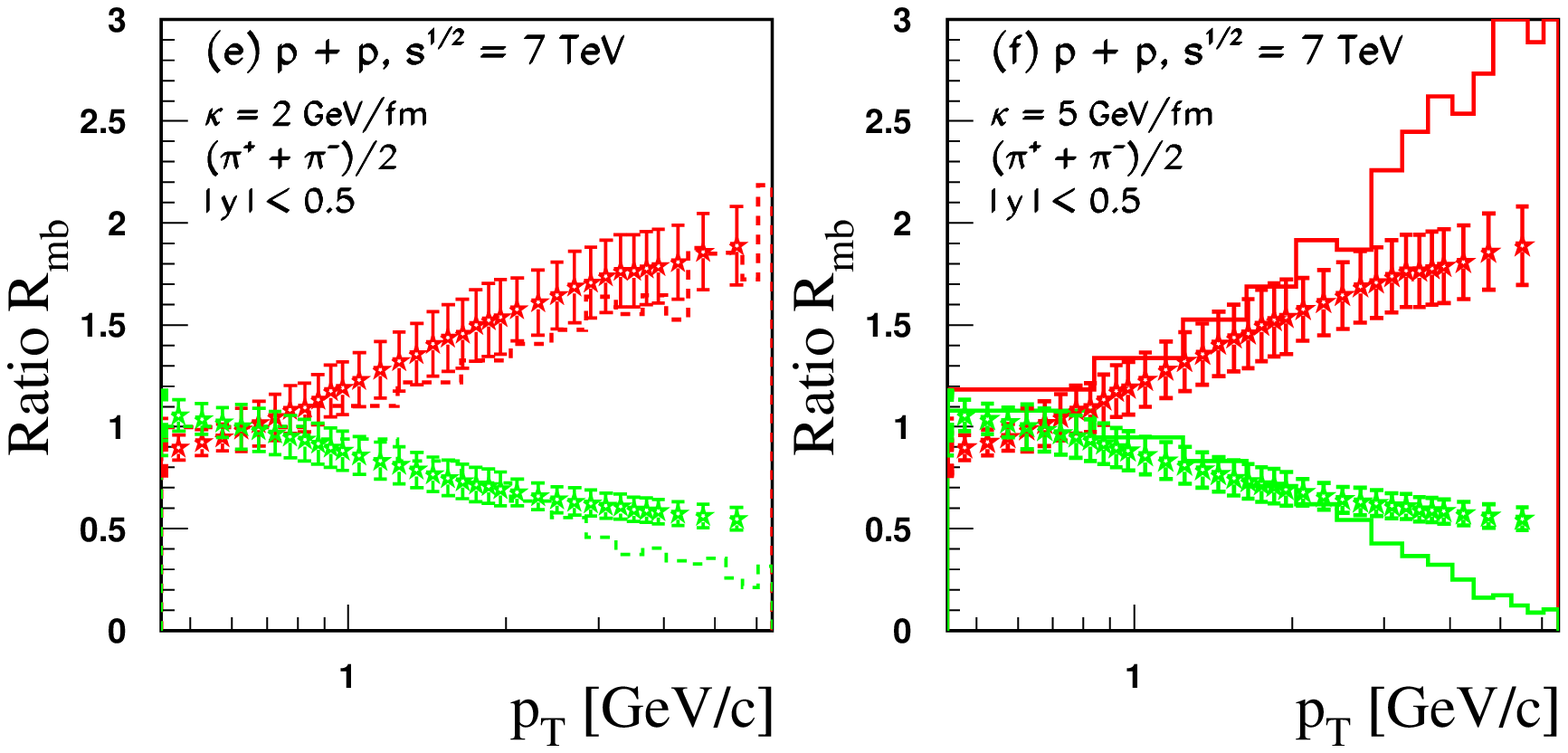}
\vskip 0.5cm
\caption[pp 7tev rmb pt pip,pim HM, LM ] 
{\small {\small HIJING/B\=B v2.0} results for
transverse momentum ($p_T$) distributions at mid-rapidity for charged 
pions in two multiplicity classes (see text for explanation). The results for 
High (Low) multiplicity class of events are presented 
in the panels a,b and panels c,d, respectively. 
The solid (dashed) histograms are obtained using  $\kappa$ = 5 GeV/fm 
($\kappa$ = 2 GeV/fm) for High (class I) and Low (class V)
multiplicity class of events.
The results for minimum bias $pp$ collisions obtained for  $\kappa$ = 2 GeV/fm
(dotted histograms) are included and compared to
data from ALICE ~\cite{Adam:2015qaa} (open circles) and CMS Collaborations~\cite{Chatrchyan:2012qb} (open squares). 
Panel e (f) include the ratios R$_{mb}$ obtained using $\kappa$ = 2 GeV/fm
($\kappa$ = 5 GeV/fm) respectively. The upper dashed and solid
histograms are for HM (class I),
and the lower dashed and solid histograms are for LM (class V) class of
events.
The experimental ratio R$_{mb}$ was calculated by us based on
average $p_T$ spectra of particle and anti-particle measured
(open stars) by ALICE
Collaboration~\cite{DerradideSouza:2016kfn,Vislavicius:2017lfr,Longpaper}.
Only statistical error bars are shown.
\label{fig:pi_hm_lm_pt_rmb}
}
\end{figure*}

For theoretical calculations within  {\small HIJING/B\=B v2.0} model 
we will chose  different classes of event activity cutting on the total 
multiplicity ($N_{ch}$) for each $10^6$ set of events generated using two effective string tension
values, {\it i.e.}, $\kappa$ = 2.0 GeV/fm and an enhanced value of $\kappa$ = 5.0 GeV/fm.
Moreover, the average charged particle density, is estimated 
(for both set of events) within each multiplicity class of events, by the integrated value of $dN_{ch}^{th}/d\eta$ at mid-pseudorapidity ($|\eta| < 0.5$). 
In addition, we generate also $10^6$ minimum bias (MB) events 
for  $\kappa$ = 2.0 GeV/fm.  
Note, that for this selection theoretical calculations 
give an integrated charged particle density at mid-pseudorapidity ($|\eta| < 0.5$), $(dN_{ch}^{th}/d\eta)^{MB}$ = 5.7, close to the experimental value 
$<dN_{ch}/d\eta> ^{MB}\,\approx\, 6.0$ quoted above.

We will consider six classes of event activity defined as:

\begin{itemize}
\item{class I: $ \,\,200 \le N_{ch} < 300$ ; $dN_{ch}^{th}/d\eta = 30.9 $ (for $\kappa$ = 2 GeV/fm);  $dN_{ch}^{th}/d\eta = 25.2$ (for  $\kappa$ = 5 GeV/fm).}          
\item{class II: $\,\,120 \le N_{ch} < 200$ ; $dN_{ch}^{th}/d\eta = 18.6 $ (for $\kappa$ = 2 GeV/fm);  $dN_{ch}^{th}/d\eta = 15.1$ (for  $\kappa$ = 5 GeV/fm).}
\item{class III:  $ \,\,100 \le N_{ch} < 120$ ; $dN_{ch}^{th}/d\eta = 12.5 $ (for $\kappa$ = 2 GeV/fm);  $dN_{ch}^{th}/d\eta = 10.3$ (for  $\kappa$ = 5 GeV/fm).} 
\item{class IV:$ \,\,80 \le N_{ch} < 100$ ; $dN_{ch}^{th}/d\eta = 9.7$ (for $\kappa$ = 2 GeV/fm);  $dN_{ch}^{th}/d\eta = 7.8$ (for  $\kappa$ = 5 GeV/fm).}
\item{class V: $ \,\,60 \le N_{ch} < 80$ ; $dN_{ch}^{th}/d\eta = 7.1 $ (for $\kappa$ = 2 GeV/fm);  $dN_{ch}^{th}/d\eta = 5.7$ (for  $\kappa$ = 5 GeV/fm).}
\item{class VI: $ \,\,30 \le N_{ch} < 60$ ; $dN_{ch}^{th}/d\eta = 4.7$ (for $\kappa$ = 2 GeV/fm);  $dN_{ch}^{th}/d\eta = 3.9$ (for  $\kappa$ = 5 GeV/fm).}
\end{itemize}

\begin{figure*} [thb]
\centering
\includegraphics[width=8.3cm,height=8.3cm]{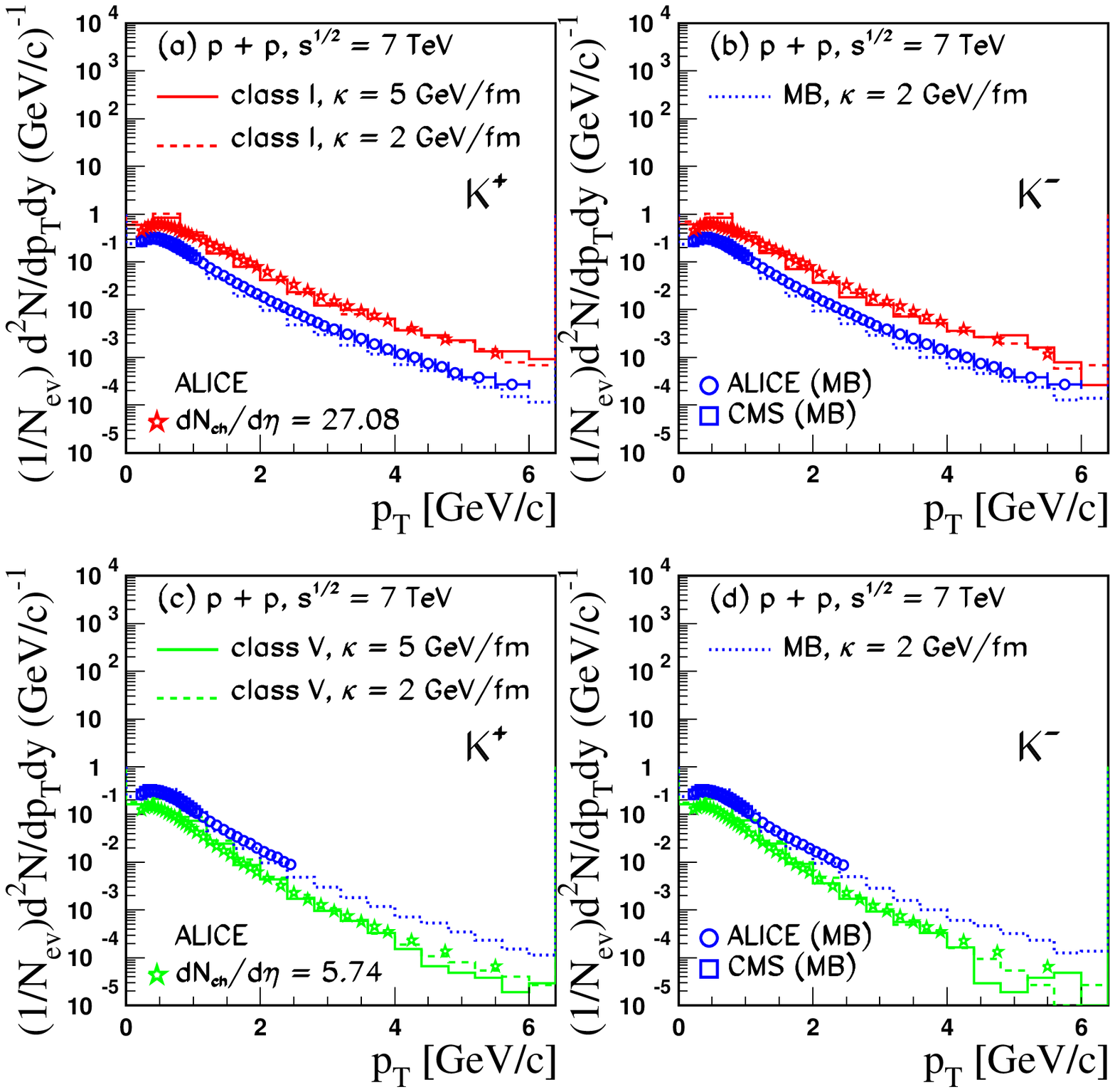}
\includegraphics[width=0.5\textwidth,height=4.0cm]{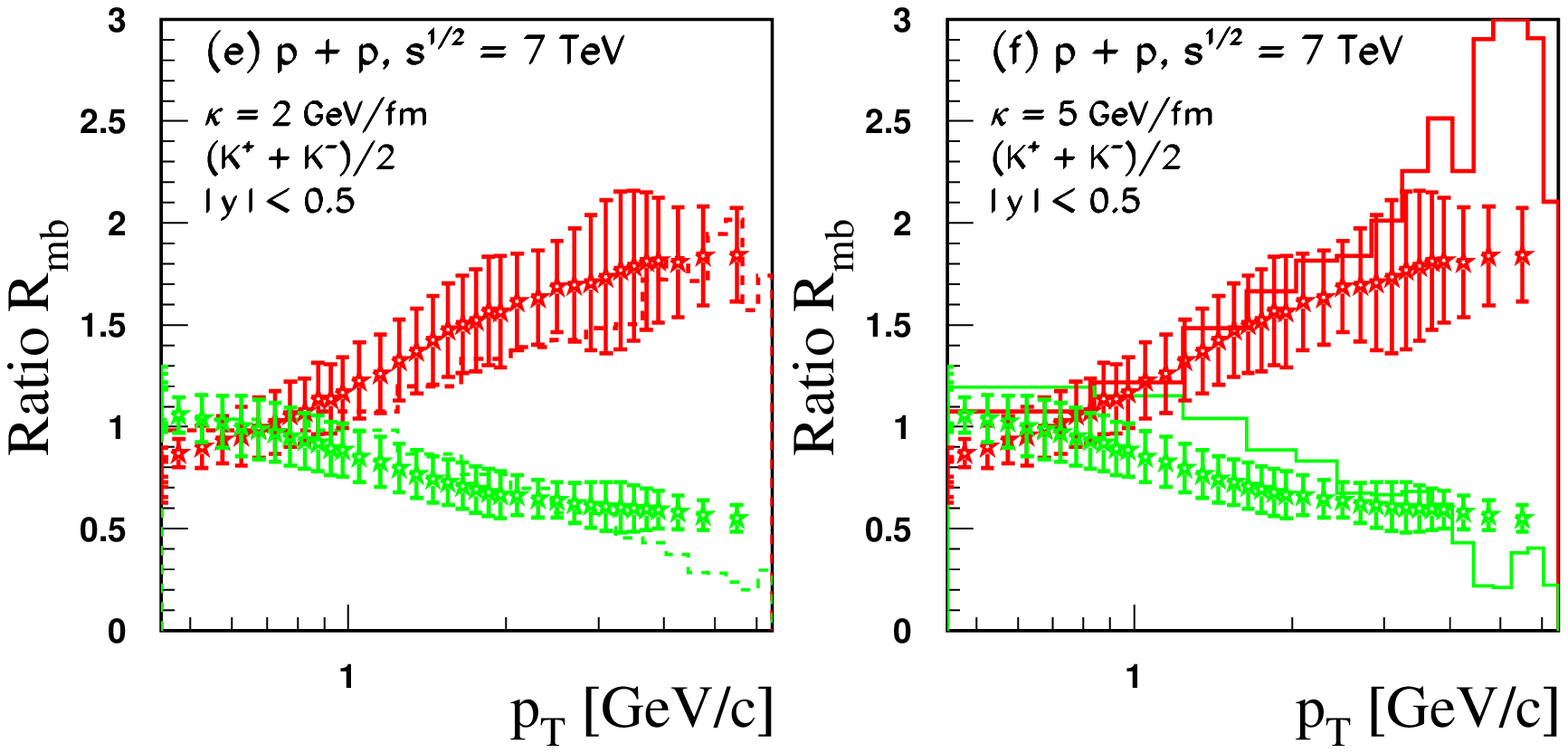}
\vskip 0.5cm
\caption[pp 7tev rmb pt kp,km HM, LM ] 
{\small The same as in Fig.~\ref{fig:pi_hm_lm_pt_rmb} for charged
kaons.
\label{fig:k_hm_lm_pt_rmb}
}
\end{figure*}
\begin{figure*} [th]
\centering
\includegraphics[width=8.3cm,height=8.3 cm]{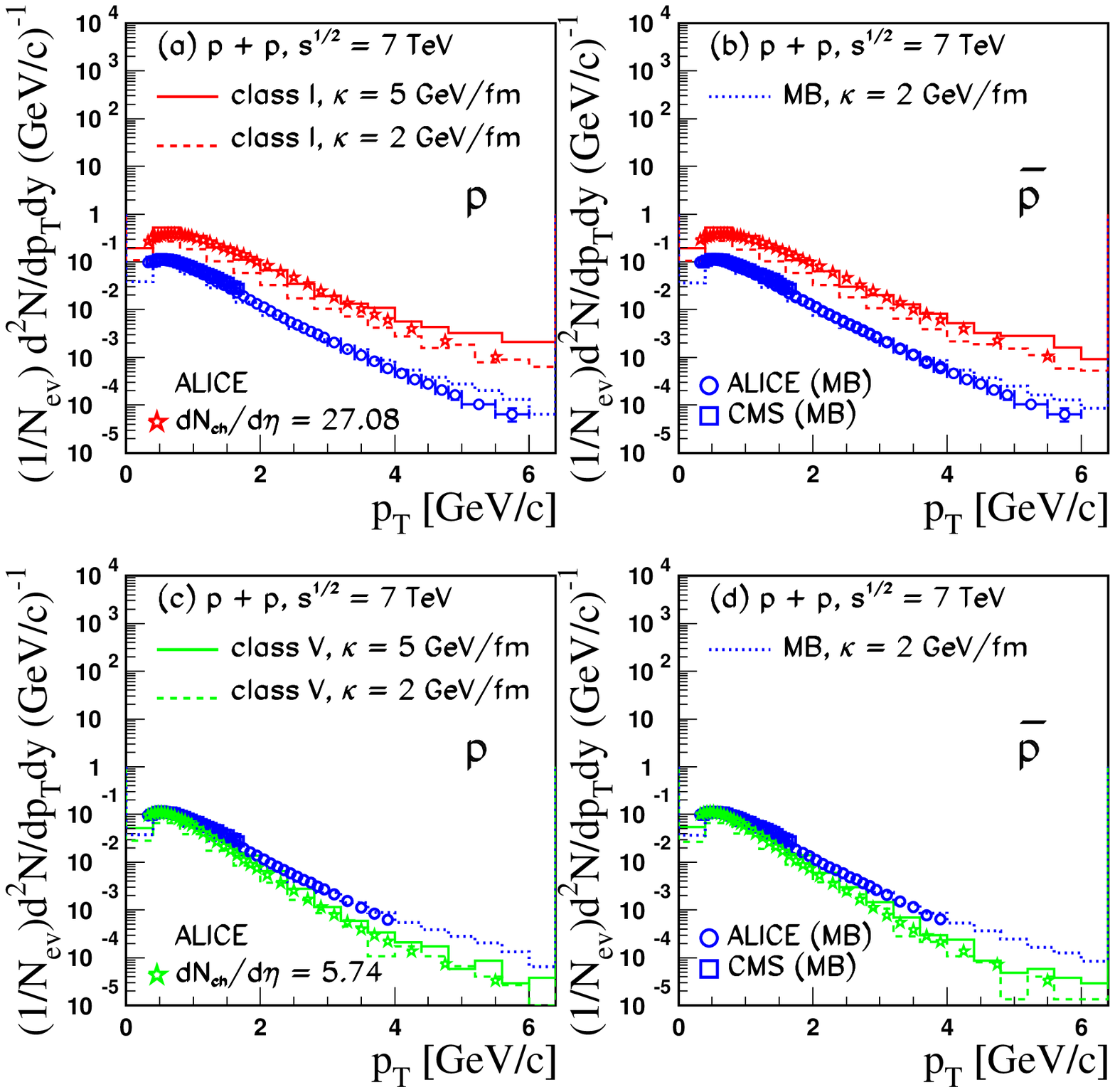}
\includegraphics[width=0.5\textwidth,height=4.0cm]{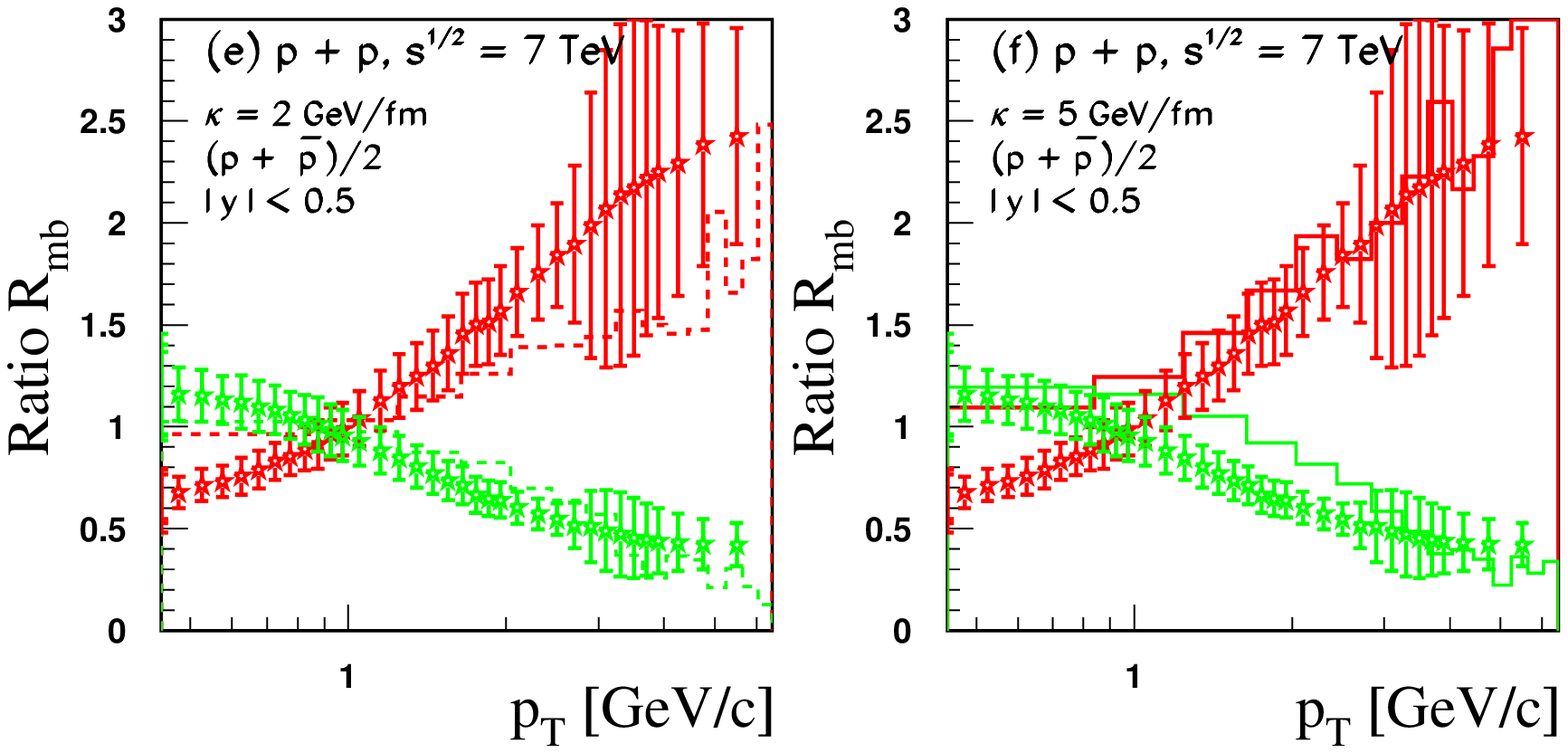}
\vskip 0.5cm
\caption[pp 7tev rmb pt p,pbar HM, LM ] 
{\small The same as in Fig.~\ref{fig:pi_hm_lm_pt_rmb} for
protons and anti-protons.
\label{fig:p_hm_lm_pt_rmb}
}
\end{figure*}

For comparison to data from Refs.~\cite{DerradideSouza:2016kfn,Vislavicius:2017lfr} we show in Fig.~\ref{fig:pi_hm_lm_pt_rmb}, Fig.~\ref{fig:k_hm_lm_pt_rmb}, and Fig.~\ref{fig:p_hm_lm_pt_rmb}  
the results of the {\small HIJING/B\=B v2.0} predictions for 
transverse momentum  distributions at mid-rapidity for 
light hadrons, {\it i.e.}, $\pi, K, p$ and their anti-particles in two multiplicity classes, class I (panels a, b) 
and class V (panels c, d). The model estimates are represented by solid (dashed) histograms for $\kappa$ = 2 GeV/fm and $\kappa$ = 5 GeV/fm, respectively.
For comparison with data, the experimental spectra (open stars) 
are chosen 
for an average value of $<dN_{ch}^{exp}/d\eta>$ similar with those 
obtained in the model, $dN_{ch}^{th}/d\eta$  (see the above six classes of event activity). 
The results for minimum bias $pp$ collisions obtained for $\kappa$ = 2 GeV/fm
are represented by dotted histograms.
Data for MB are from Ref.~\cite{Adam:2015qaa} (open circles) and Ref.~\cite{Chatrchyan:2012qb} (open squares).

\begin{figure*} [th]
\centering
\includegraphics[width=0.8\linewidth,height=5.0cm]{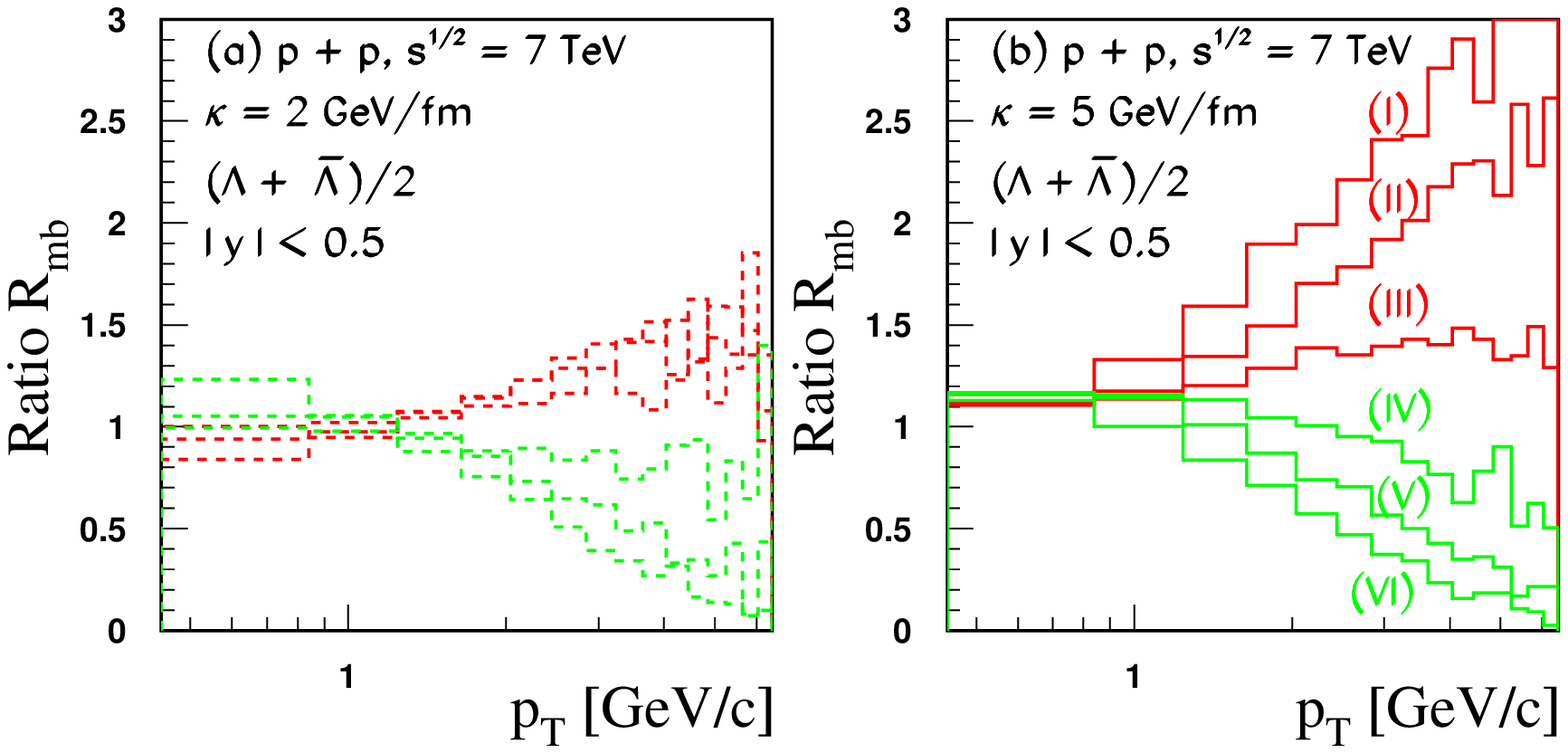}
\vskip 0.5cm\caption[Rmb ratio l0 all HM,LM ks2 ks5] 
{\small The  {\small HIJING/B\=B v2.0} model predictions for
$\Lambda$ + $\bar{\Lambda}$ produced in $pp$ collisions 
at $\sqrt{s}$ = 7 TeV. 
The ratios of the normalized $p_T$ distributions, $R_{mb}$ (see Eq.~\ref{eq:rmb}) for six multiplicity classes (see text for explanation) based on 
average $p_T$ spectra of particle and anti-particle.
 From top to bottom the calculations correspond 
to class (I) to class (VI) multiplicity events.  
Left-the results obtained with $\kappa$ = 2 GeV/fm; Right- the results obtained with ($\kappa$ = 5 GeV/fm).  
\label{fig:l0_allcuts_hm_lm_rmb}
}
\end{figure*}

The ratio of double differential cross sections normalized to the charged particle densities, $R_{mb}$ 
(calculated by us) is plotted  for high (class I) and low (class V) multiplicity classes in panels e, f by dashed and solid histograms
for $\kappa$ = 2 GeV/fm
and $\kappa$ = 5 GeV/fm, respectively.  
This ratio is  
based on average $p_T$ spectra of particle and anti-particle measured (open stars) by ALICE Collaboration~\cite{DerradideSouza:2016kfn,Vislavicius:2017lfr,Longpaper}. 
In the calculations we take into account the variation of strong color (electric) field with 
energy. The assumed effective value of the string tension is  
$\kappa$ = 2 GeV/fm (panel e) corresponding to
$\kappa(s)= \kappa_{0} \,\,(s/s_{0})^{0.04}\,\,{\rm GeV/fm} $ 
(see Eq.~\ref{eq:kappa_sup}).
Since we expect in high multiplicity proton-proton collisions features that are similar to those 
observed in Pb - Pb collisions,
we consider also the results obtained for an enhanced value of effective string tension 
to  $\kappa$ = 5 GeV/fm (see panel f).  
The agreement with the data is fairly good in the limit of the error bars, 
except for very low $p_T < 1 $ GeV values.
The experimental spectra show a small depletion 
at high multiplicity at very low 
$p_T$, indicating possible influence of the radial flow.
The transverse momentum spectra of identified particles carrying light quarks 
and their azimuthal distributions are well described by hydrodynamical models 
\cite{Pierog:2013ria,Werner:2013tya} at very low $p_T$.
However, as far as in the string model the pressure is not considered,
it is not expected to describe such effects which could originate from collective expansion.  
At low and intermediate ${\rm p}_{\rm T}$ ($1$ GeV/c $\,<\,{\rm p}_{\rm T}\,< 6 $ GeV/c), 
the non-perturbative production mechanism via SLCF  
produces a clear split between High and Low multiplicity events.
For the highest multiplicity (class I), we see a  hardening 
of the  ${\rm p}_{\rm T}$ spectra for $\pi, K, p$ and their anti-particles. 
However, 
within the experimental errors,
the agreement between the model predictions and experiment in terms of 
$R_{mb}$ is rather similar for both values of the string tension, i.e. 
$\kappa$ = 2 GeV/fm and $\kappa$ = 5 GeV/fm for pions and kaons.
For protons, the agreement is definitely better at low charged particle multiplicity (class V) for $\kappa$ = 2 GeV/fm and at higher charged particle multimplicity (class I) for $\kappa$ = 5 GeV/fm.
Due to strange quark content of (multi)strange particle  
the study of the ratio $R_{mb}$ is of particular interest. 
Since we expect higher sensitivity to SLCF effects for (multi)strange
than for bulk particles, measurements of $p_T$ distributions at mid-rapidity as well as the ratio $R_{mb}$ 
could help to evidentiate these effects, 
within the phenomenology embedded in {\small HIJING/B\=B v2.0} model.

Figure~\ref{fig:l0_allcuts_hm_lm_rmb} show 
the ratios of the normalized $p_T$ distributions, R$_{mb}$  
of $\Lambda +\bar{\Lambda}$ 
produced in {\it p}+{\it p} collisions at $\sqrt{s} = 7$ TeV.
The results for six multiplicity classes (class I to class VI) based on 
average $p_T$ spectra of particle and anti-particle are included.
 From top to bottom the calculations correspond 
to highest (class I) to lowest (class VI)  multiplicity events. 
Left (Right) panels are the results obtained with $\kappa$ = 2 GeV/fm ($\kappa$ = 5 GeV/fm) respectively.  
We remark a clear hardening of the $p_T$ spectra for high multiplicity, 
especially for $N_{ch} > 100$ (class III to class I events), where a change in 
the slope is obvious. 
The effect is more evident  for an enhanced  effective value of string 
tension $\kappa$ = 5 GeV/fm (see Fig.~\ref{fig:l0_allcuts_hm_lm_rmb} right panel).
Similar results (not included here) are obtained for multi-strange
particles, {\it i.e.}, $\Xi$ and $\Omega$. 
High multiplicity events have a higher fraction of heavier particles,
meaning with a higher strangeness content.
We can explain this fact as an effect of strong color field embedded in our model.
Note, that $N_{ch} > 120$ is also the charged particle multiplicity above which
was observed the enhancement in the near side long range two-particle
correlation reported by CMS Collaboration ~\cite{Khachatryan:2010gv}.
However, there is no mechanism that produces a ridge in our model.

\begin{figure*} [t!]
\centering
\includegraphics[width=8.3cm,height=8.3 cm]{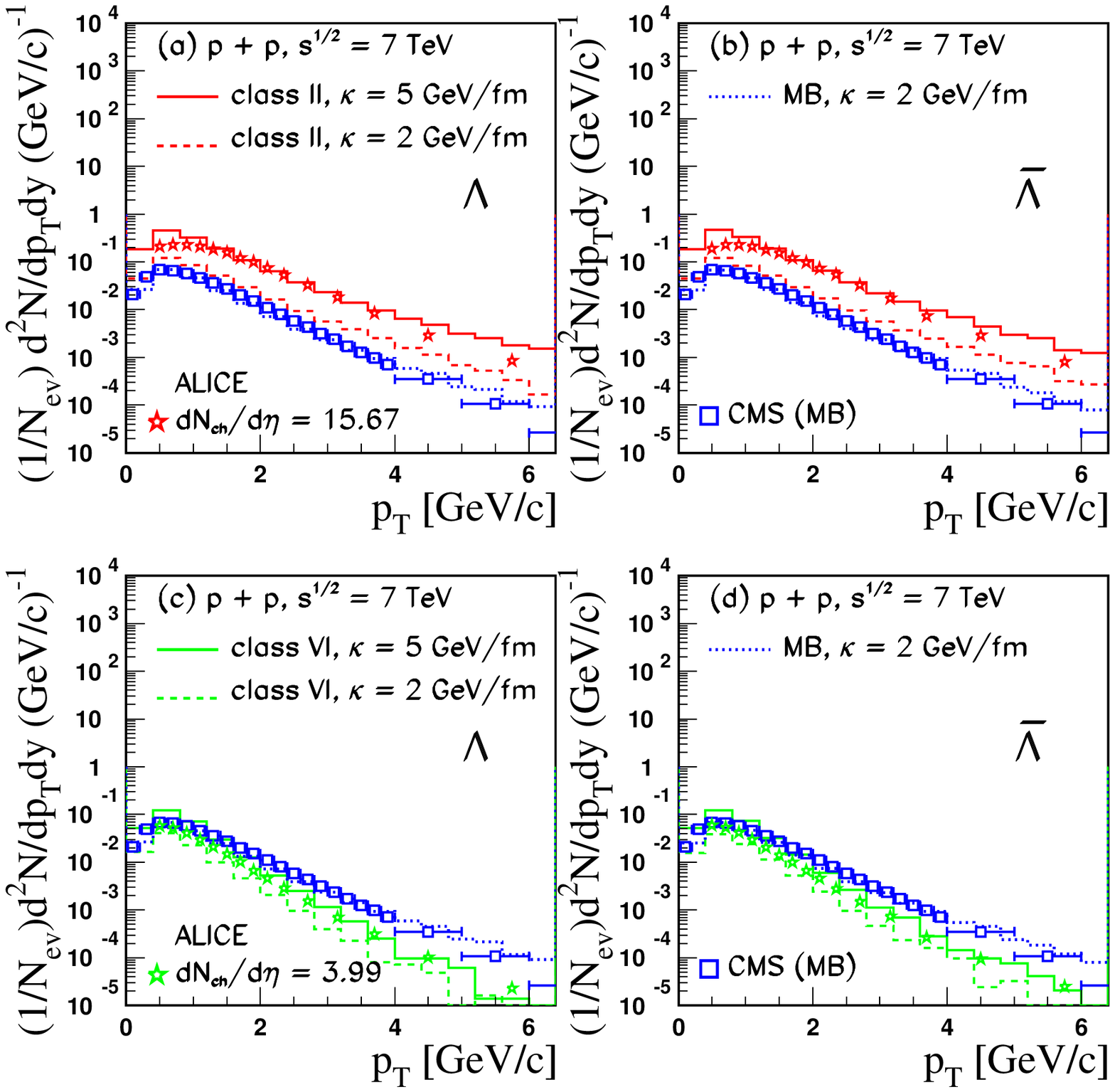}
\includegraphics[width=0.5\textwidth,height=4.0cm]{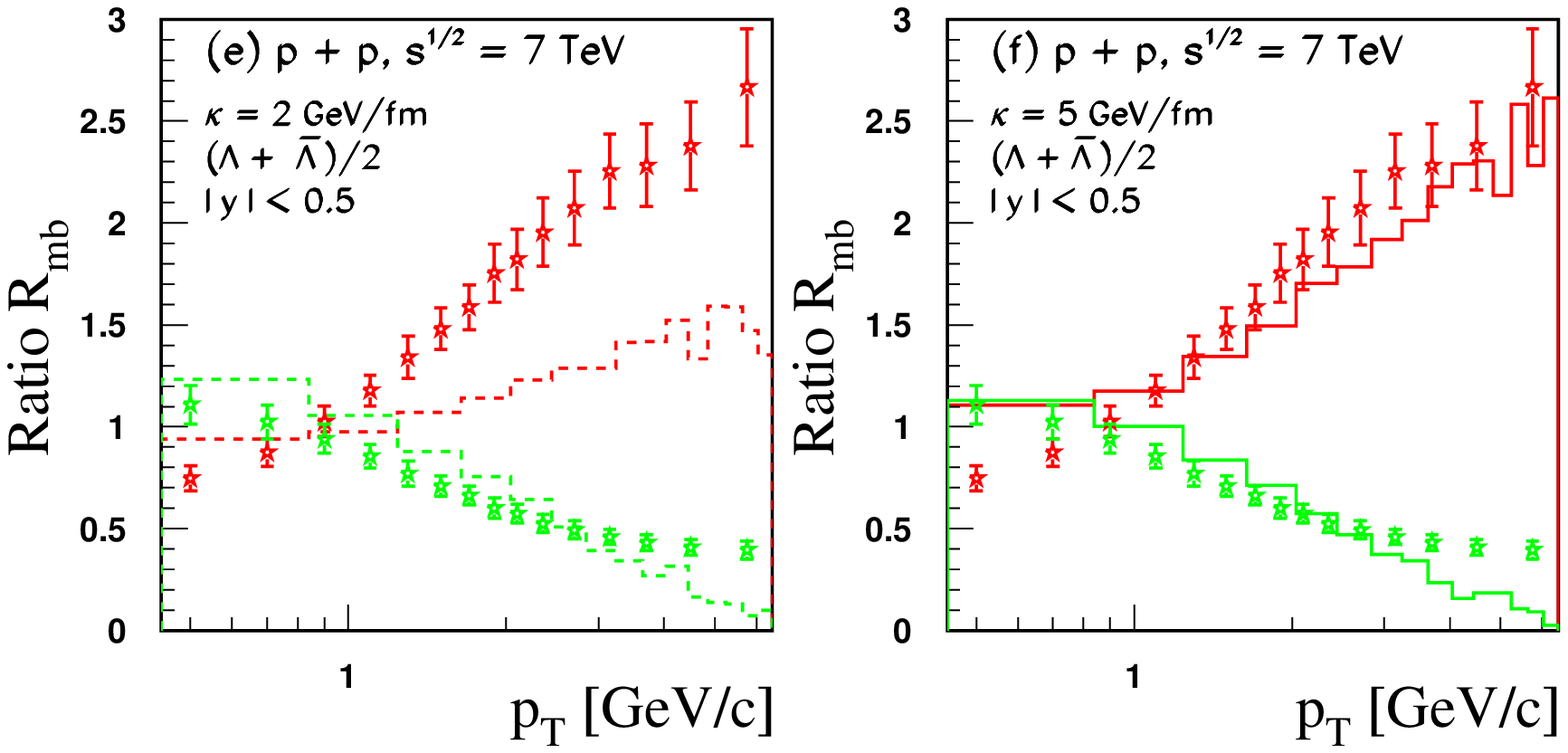}
\vskip 0.5cm
\caption[pp 7tev rmb pt l0,l0bar HM, LM ] 
{\small The same as in Fig.~\ref{fig:pi_hm_lm_pt_rmb} for 
events of high (class II) and low (class VI) multiplicity. 
The calculations are for $\Lambda$ and $\bar{\Lambda}$. 
The results for minimum bias $pp$ collisions obtained with $\kappa$ = 2 GeV/fm
(dotted histograms) are included and compared to
data from CMS Collaborations~\cite{Chatrchyan:2012qb} (open squares).
Only statistical error bars are shown. 
\label{fig:l0_hm_lm_pt_rmb}
}
\end{figure*}

\begin{figure*} [th!]
\centering
\includegraphics[width=8.3cm,height=8.3 cm]{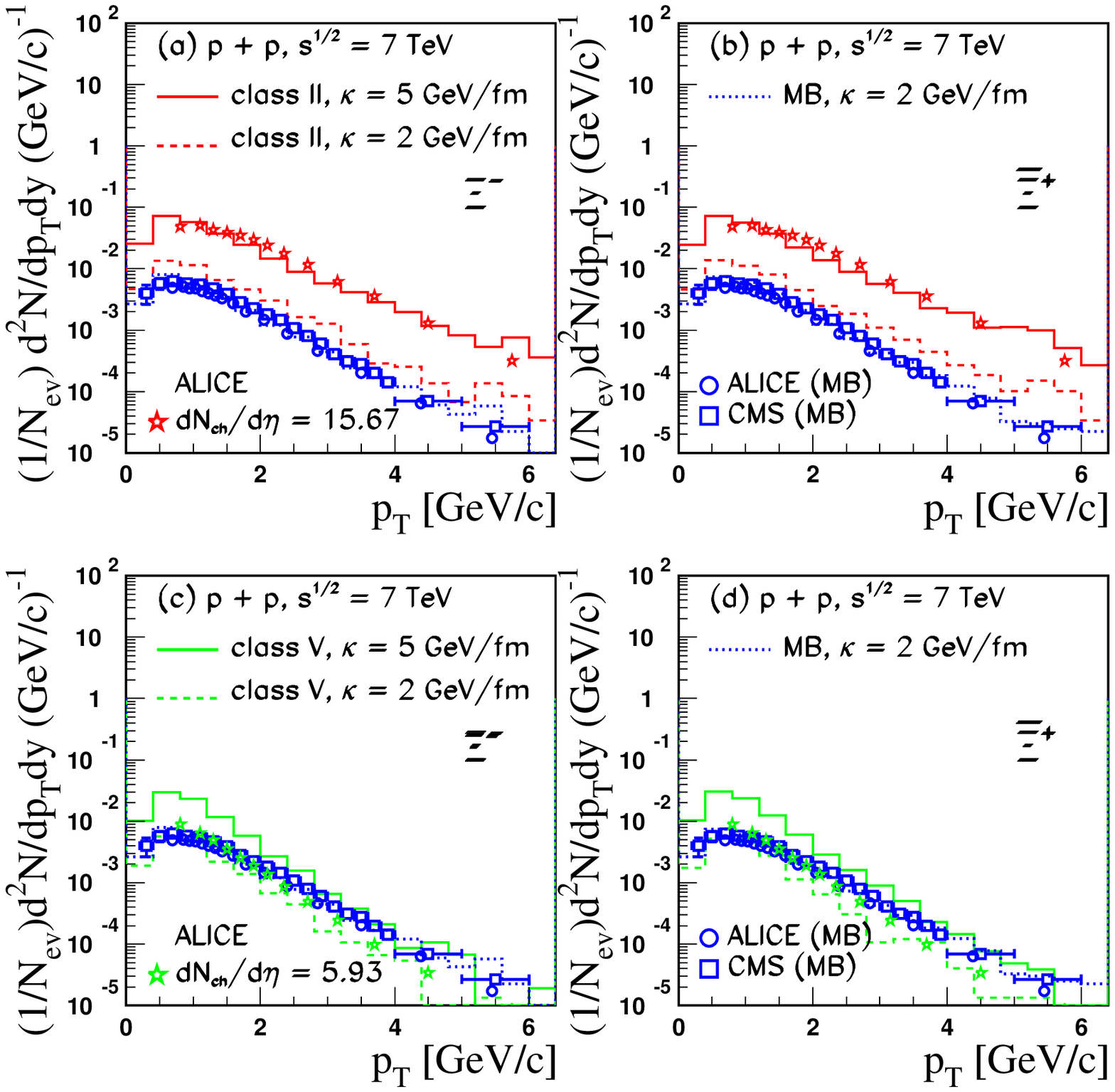}
\includegraphics[width=0.5\textwidth,height=4.0cm]{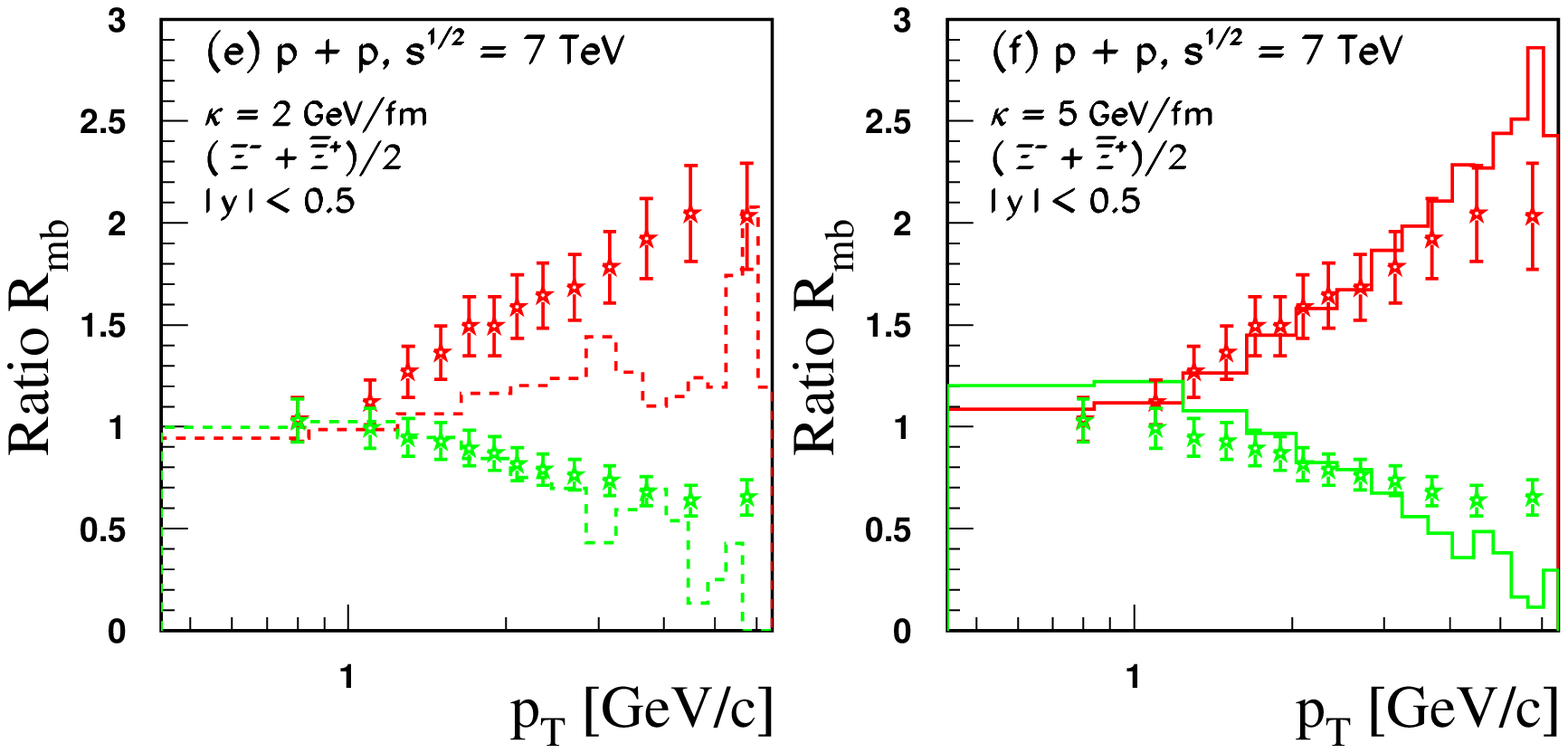}
\vskip 0.5cm
\caption[pp 7tev rmb pt xi,xibar HM, LM ] 
{\small The same as Fig.\ref{fig:pi_hm_lm_pt_rmb} for 
events of high (class II) and low (class V) multiplicity. 
The calculations are for for $\Xi^-$ and ${\bar{\Xi}}^+$.
The results for minimum bias $pp$ collisions obtained with $\kappa$ = 2 GeV/fm
(dotted histograms) are included and compared to
data from ALICE ~\cite{Bianchi:2016szl,DerradideSouza:2016kfn,Longpaper} (open circles) 
and CMS Collaborations~\cite{Chatrchyan:2012qb} (open squares). 
Only statistical error bars are shown.
\label{fig:xi_hm_lm_pt_rmb}
}
\end{figure*}

\begin{figure*} [th]
\centering
\includegraphics[width=8.3cm,height=8.3 cm]{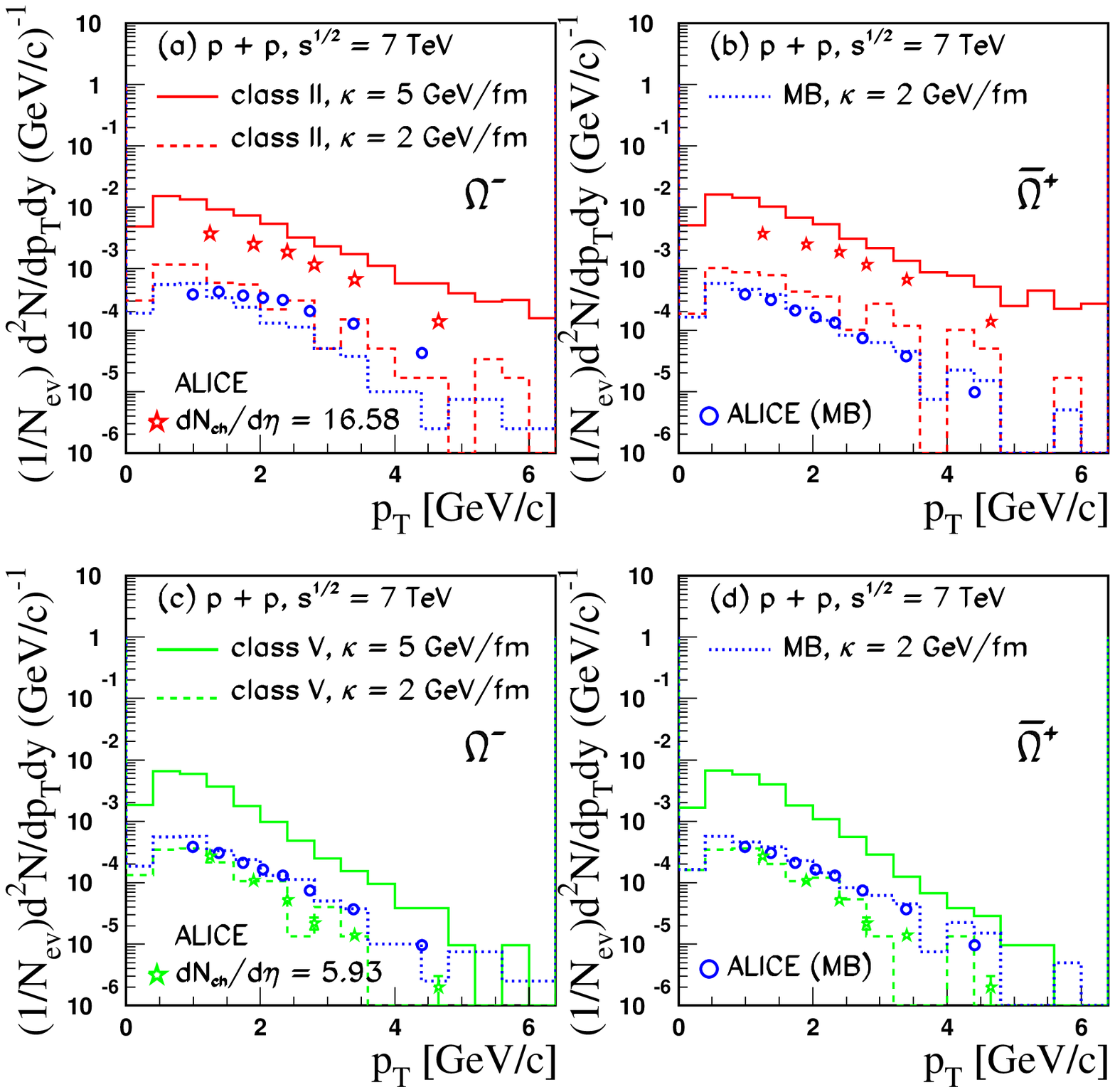}
\includegraphics[width=0.5\textwidth,height=4.0cm]{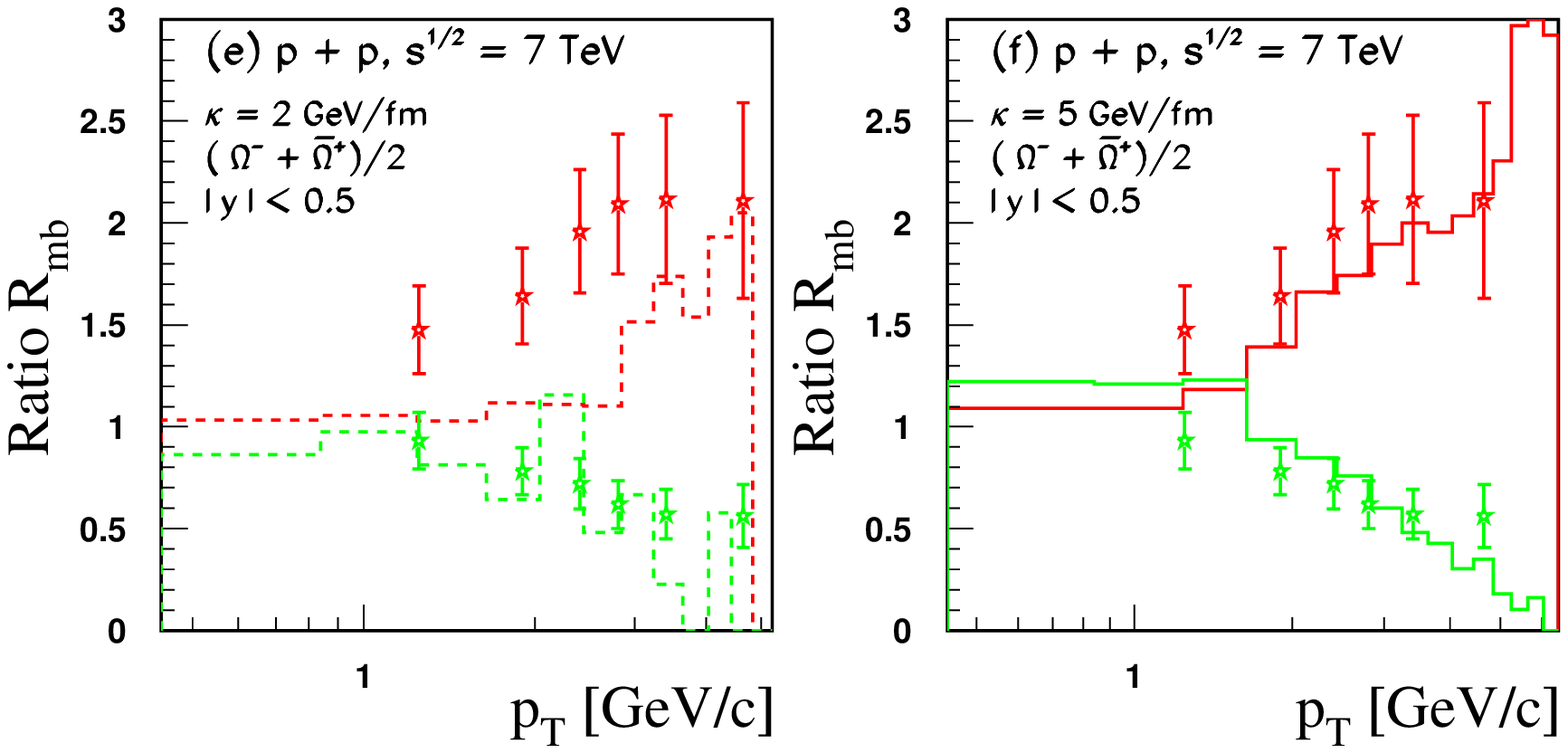}
\vskip 0.5cm
\caption[pp 7tev rmb pt omm,omp HM, LM ] 
{\small The same as in Fig.~\ref{fig:pi_hm_lm_pt_rmb} for 
events of high( class II) and low (class V) multiplicity. The calculations are
 for $\Omega^-$ and ${\bar{\Omega}}^+$.
The results for minimum bias $pp$ collisions obtained with $\kappa$ = 2 GeV/fm
(dotted histograms) are included and compared to
data (open circles) from ALICE ~\cite{Bianchi:2016szl,DerradideSouza:2016kfn,Longpaper}. 
Only statistical error bars are shown.
\label{fig:om_hm_lm_pt_rmb}
}
\end{figure*}

 The experimental fact that $pp$ collisions manifest features similar with 
Pb - Pb collisions  \cite{Khachatryan:2010gv,CMS:2012qk,Abelev:2012ola,Aad:2013fja, Andrei:2014vaa,Aad:2015gqa,Witek:2017dyn,ALICE:2017jyt,Petrovici:2018mpq}
 point out to the 
necessity to modify $\kappa$ , in describing observables in $pp$ collisions for HM class of events.
The calculations with SLCF contributions assume 
an effective string tension value  
$\kappa$ =  2 GeV/fm, obtained from an energy depend  $\kappa$ (see Sec. II), 
while the results with $\kappa$ = 5 GeV/fm
are obtained based on the above experimental fact.
Note, that a specific size dependent $\kappa$ = $\kappa(r)$ 
was considered recently in PYTHIA 8 model, with r 
a new parameter fixed to fit data~\cite{Fischer:2016zzs}.

Therefore, in 
Fig.~\ref{fig:l0_hm_lm_pt_rmb} ($\Lambda$ and $\bar{\Lambda}$)
, Fig.~\ref{fig:xi_hm_lm_pt_rmb} ($\Xi^-$ and ${\bar{\Xi}}^+$), and
Fig.~\ref{fig:om_hm_lm_pt_rmb} ($\Omega^-$ and ${\bar{\Omega}}^+$) we 
show the results obtained for $p_T$ distributions at mid-rapidity 
for (multi)strange particles in two event classes,
corresponding to high (HM) and low (LM) charged particle multiplicity, 
The calculations for minimum bias events are included 
and compared to data from Refs.~\cite{Bianchi:2016szl,DerradideSouza:2016kfn,Chatrchyan:2012qb}.
As in the previous calculations comparison to data for HM and LM events, 
is made for $p_T$ spectra 
obtained for a class of events which give a value of
$dN_{ch}^{th}/d\eta$  similar with those obtained in the experiment  
$<dN_{ch}^{exp}/d\eta>$ . 
Theoretical predictions for the $p_T $ dependence
of $R_{mb}$ for $\Lambda +\bar{\Lambda}$, $\Xi^- + {\bar{\Xi}}^+$, and 
$\Omega^- + \bar{\Omega}^+$ are presented for two scenarios: 
using $\kappa$ = 2 GeV/fm (panel e) and an increased value to 
$\kappa$ = 5 GeV/fm (panel f). 
 The results show a clear hardening of $p_T$ spectra in case of HM class of events.
Moreover, in case of the LM class of events,
the $R_{mb}$ ratio of (multi)strange particles are better
described using $\kappa$ = 2 GeV/fm.
In contrast, an increase of effective value $\kappa$  to  $\kappa$ = 5 GeV/fm 
better describes  class with HM events.
The remark is true for strange $\Lambda + \bar{\Lambda}$ as well as 
for multi-strange ($\Xi^- + {\bar{\Xi}}^+$, $\Omega^- + {\bar{\Omega}}^+$ ) particles 
in $pp$ collisions at $\sqrt{s} = $ 7 TeV.

To conclude, for a better description of (multi)strange particle productions    
in high charged particle multiplicity $pp$ collisions, we have to consider an increase of effective string tension value from 
$\kappa$ = 2 GeV/fm to   
$\kappa$ = 5 GeV/fm, is strongly supported by data.
The fact that an effective value $\kappa$ = 5 GeV/fm describes better 
the $R_{mb}$ ratio in $pp$ collisions at $\sqrt{s} = $ 7 TeV, 
reveal features similar with those observed in chromoelectric flux 
configurations used to describe some experimental observables in 
Pb - Pb collisions at $\sqrt{s_{NN}} = $ 2.76 TeV \cite{Pop:2013ypa}.
The enhancement of (multi)strange hadron yields 
as function of multiplicity have been associated with the creation of a 
strongly interacting medium, sQGP \cite{Abelev:2013haa}.
Recently, a similar behavior was also observed for multi-strange hadrons in 
high-multiplicity $pp$ collisions \cite{ALICE:2017jyt} and this observation
challenge all string fragmentation models \cite{Fischer:2016zzs}.
Finally, we remark that for $pp$ collisions at $\sqrt{s}$ = 7 TeV 
our model predicts  
higher sensitivity to SLCF effects for ID (multi)strange ($\Lambda$, $\Xi$,
$\Omega$) than for light hadrons ($\pi, K, p$).
The calculations assuming an effective  string tension value 
which vary only with 
energy as $\kappa(s)= \kappa_{0} \,\,(s/s_{0})^{0.04}\,\,{\rm GeV/fm}$ describe 
fairly well the (multi)strangeness production in the LM event classes,
but fail to describe (multi)strange production in HM event classes.
A better description is obtained for an enhanced effective string tension value
$\kappa$ = 5 GeV/fm which point out to the necessity of a new dependency on multiplicity 
(or  $\epsilon_{\rm ini}$ ) for the effective string tension value. 

\section{Summary and Conclusions}

In summary, we studied in the framework of the {\small HIJING/B\=B v2.0} model,
the influence of possible strong homogeneous 
constant color electric fields on new experimental observables measured 
by ALICE Collaboration, especially for identified particle 
in  $pp$, $p$ - Pb, and  Pb - Pb collisions at  $\sqrt{s}$ = 7 TeV, 
$\sqrt{s_{\rm NN}}$ = 5.02 TeV, and   $\sqrt{s_{\rm NN}}$ = 2.76 TeV,
respectively.
The effective string tension $\kappa$, control Q\=Q pair creation rates 
and the suppression factors $\gamma_{Q\bar{Q}}$. 
The measured  average transverse momentum  and  ratio 
$R_{mb}$ of ID particle help to verify our assumptions 
and to set the strangeness suppression factor.  
We assume in our calculations 
an energy and possible system dependence of the effective string tension,
$\kappa$.

For Pb - Pb collisions at $\sqrt{s_{\rm NN}}$ = 2.76 TeV  
all nuclear effects included in the model, {\it e.g.}, 
strong color fields, shadowing and quenching should be taken into account.
However, partonic energy loss and jet quenching process, 
as embedded in the model, brought a fair description of the 
$p_T$ distributions of identified light hadrons ($\pi, K, p$).
The discrepancy could be explained by 
an initial condition with a large pressure and therefore 
a large collective flow, 
which is not embedded in our model.

For identified particle in  $pp$ collisions at  $\sqrt{s}$ = 7 TeV
we compute correlation between mean 
transverse momentum and multiplicity of charged particles (N$_{ch}^*$) 
at central rapidity as well as 
the ratio of double differential cross sections 
normalized to the charged particle densities versus multiplicity, $R_{mb}$.
In the calculations we take into account the variation of strong color 
(electric) field with energy but not with the multiplicity (or initial energy densities, $\epsilon_{\rm ini}$) 
of the colliding system. The assumed effective string tension is  
$\kappa$ = 2 GeV/fm, corresponding to
$\kappa(s)= \kappa_{0} \,\,(s/s_{0})^{0.04}\,\,{\rm GeV/fm} $ (see Eq.~\ref{eq:kappa_sup}).
Since we expect in high multiplicity proton-proton collisions features that are similar 
to those observed in Pb - Pb collisions,
we consider also the results obtained with an enhanced value of the effective string tension, 
from $\kappa$ = 2 GeV/fm to  $\kappa$ = 5 GeV/fm. 
This increase of the strength of color fields lead to 
a ratio  $R_{mb}$ consistent with recent data for HM class of events, 
while the LM class of events are
better described using  a lower effective string tension 
value $\kappa$ = 2 GeV/fm.
These results show that the above increase 
of the strength of color fields could be  
an important dynamical mechanisms. 
New measurements with high statistics at low and intermediate $p_T$ 
 ($1 < p_T < 6 $ GeV/{\it c}) of the ratio  $R_{mb}$
in $pp$ collisions at LHC energies, could help to disentangle  
between different model approaches and/or different dynamical mechanisms,
especially for high multiplicity event classes. 

Note, that the {\small HIJING/B\=B model} is based on a time-independent 
strength of color field, while in 
reality the production of Q\=Q pairs is 
a time and space dependent phenomenon,
being far from equilibrium. To achieve more quantitative
conclusions, such time and space dependent mechanisms 
\cite{Hebenstreit:2008ae,Levai:2009mn}
should be considered in the next generation of Monte Carlo codes.
\vspace{0.5cm} 
\section{Acknowledgments}
\vskip 0.2cm 
One of us (VTP) would like to acknowledge useful discussions with Miklos Gyulassy and Jean Barrette in the early phase of analysis.  
This work was supported by the Natural Sciences and Engineering 
Research Council of Canada, and by the Projects number 44/05.10.2011 
and 4/16.03.2016 of the Ministry of Research and
Innovation via CNCSI and IFA coordinating agencies.

\end{document}